\documentclass[12pt, english, openany]{book}

\usepackage{cite}
\usepackage{algorithmic}
\usepackage{graphicx, caption}
\usepackage{textcomp}
\usepackage{xcolor}
\usepackage{lineno}
\usepackage{wrapfig}
\PassOptionsToPackage{hyphens}{url}
\usepackage{hyperref}
\usepackage{enumitem}
\usepackage[
    type={CC},
    modifier={by},
    version={4.0},
]{doclicense}
\usepackage[utf8]{inputenc}
\usepackage{optidef}
\usepackage{multimedia}
\usepackage{amssymb}
\usepackage{multicol}
\usepackage{soul}
\usepackage{subcaption}
\usepackage[]{graphicx}
\usepackage[]{color}
\usepackage{alltt}
\usepackage[T1]{fontenc}
\usepackage[utf8]{inputenc}
\usepackage[parfill]{parskip}
\usepackage{pgffor}
\usepackage{setspace}
\usepackage{datetime}
\usepackage{wrapfig}
\usepackage[top=100pt,bottom=100pt,left=68pt,right=66pt]{geometry}
\usepackage{lipsum}
\usepackage{pdfpages}
\usepackage{tabulary}
\usepackage{multicol}

\usepackage[section]{placeins}
\usepackage{afterpage}

\usepackage[english, italian]{babel}

\usepackage{fancyhdr}
\usepackage{titlesec}
\titleformat{\chapter}
   {\normalfont\LARGE\bfseries}{\thechapter.}{1em}{}
\titlespacing{\chapter}{0pt}{50pt}{2\baselineskip}

\usepackage{float}
\floatstyle{plaintop}
\restylefloat{table}

\usepackage[tableposition=top]{caption}
\usepackage{hyperref}
\hypersetup{hidelinks,linkcolor = black}

\graphicspath{ {figures/} }
\frontmatter

\begin{document}

\selectlanguage{english}

\begin{titlepage}
	\clearpage\thispagestyle{empty}
	\centering
	\vspace{1cm}

	{\normalsize \textit{Neither a person nor an apple can be diverse. Diversity is the property of a collection of people—a basket with many kinds of fruit. \\
		– Scott E. Page} \par}
		
	\vspace{1.2cm}
	
	{\Huge \textbf{AI Governance and Ethics Framework for \\\vspace{0.5cm} Sustainable AI and Sustainability}} \\

	\vspace{1.2cm}
    
    \centering \includegraphics[width=0.9\textwidth]{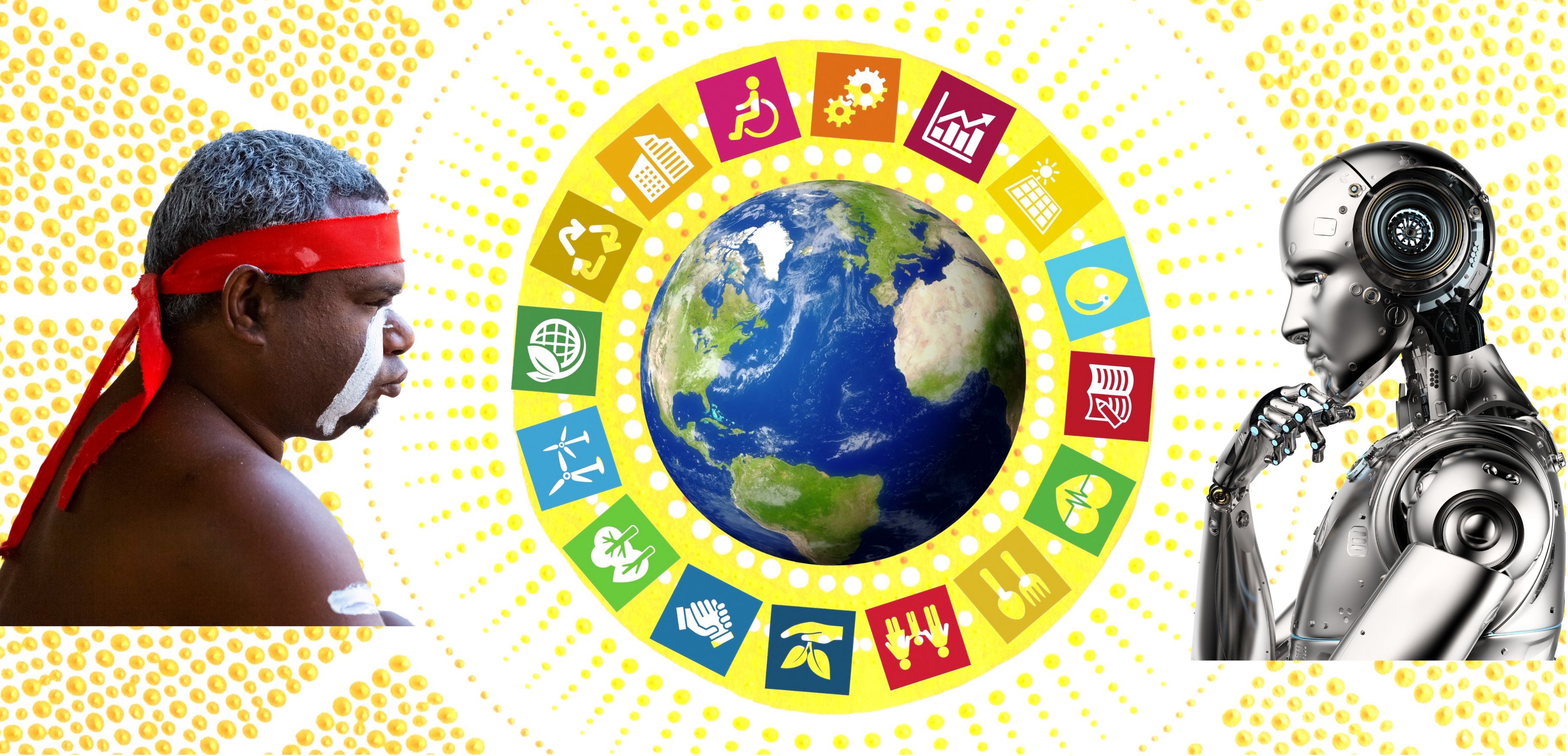}
    
    \vspace{0.6cm}
    
    \text{Dr Mahendra Samarawickrama (GAICD, MBA, SMIEEE, ACS(CP))}
    
    \vfill
    
	{\normalsize May 18, 2022 \par}
	
	\begin{figure}[b]
	\centering
    \href{https://bit.ly/AIESG}
    {\includegraphics[width=0.15\textwidth]{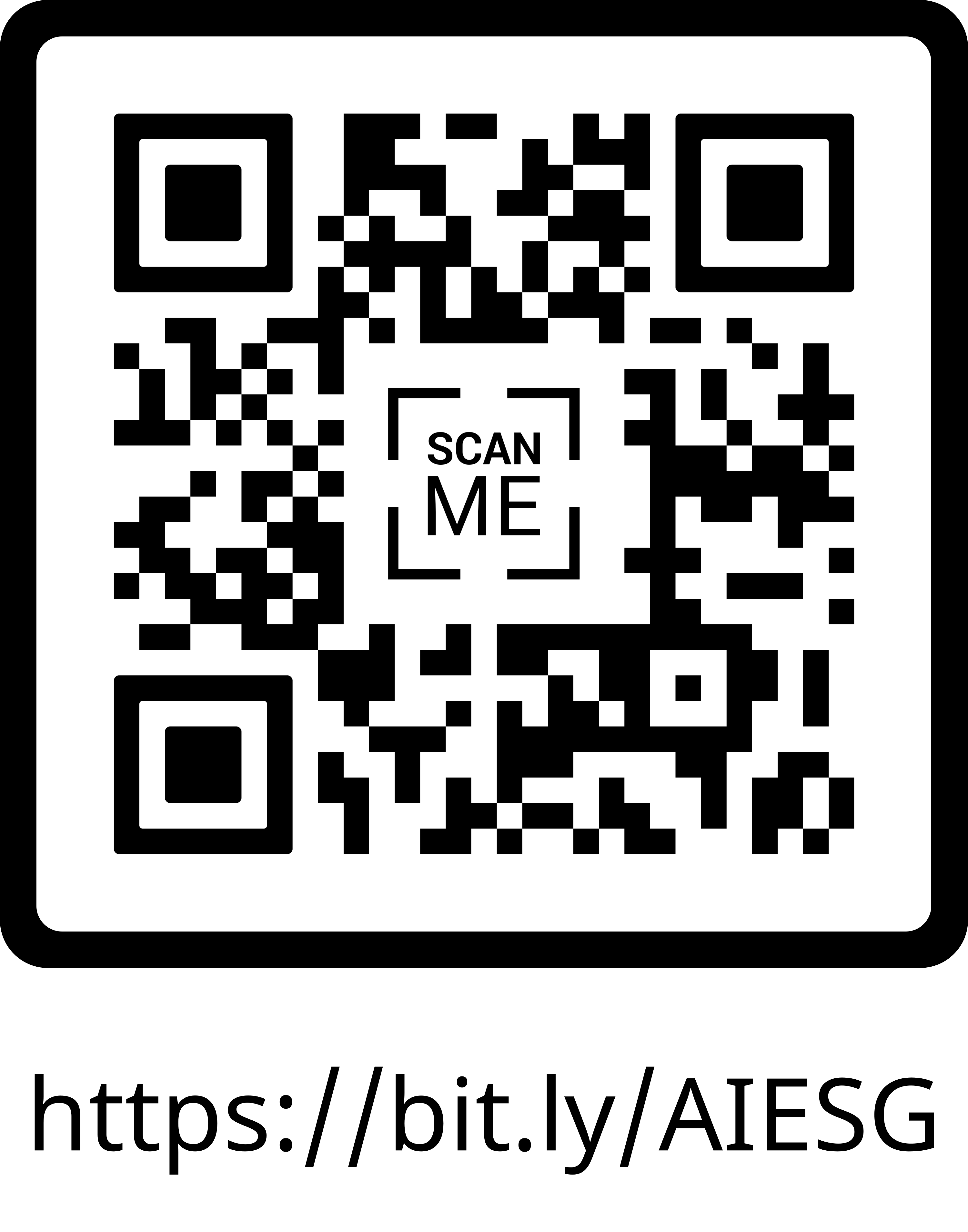}}
    \end{figure}
    
	\pagebreak

\end{titlepage}

\pagebreak
\chapter*{Copyright Notice}

Copyright © 2022 Mahendra Samarawickrama
\newline \newline
ISBN: 978-0-6454693-0-1

\doclicenseThis
\noindent
This report was submitted to the consultation process of The Australian Department of the Prime Minister and Cabinet for the regulation of artificial intelligence (AI) and automated decision making.

\subsection*{Third party copyright} 
Wherever a third party holds copyright in this material, the copyright remains with that party. Their permission may be required to use the material. Please contact them directly.

\subsection*{Attribution}  
This publication should be attributed as follows:

M. Samarawickrama, “AI Governance and Ethics Framework for Sustainable AI and Sustainability,” \textit{Submission in response to the Department of the Prime Minister and Cabinet issues paper Positioning Australia as a leader in digital economy regulation - Automated Decision Making and AI Regulation}, Apr. 2022, ISBN: 978-0-6454693-0-1.

\pagebreak

\thispagestyle{plain}
\begin{center}
    \Large
    \textbf{AI Governance and Ethics Framework for Sustainable AI and Sustainability}
        
    \vspace{0.4cm}
    {\normalsize Dr Mahendra Samarawickrama (GAICD, MBA, SMIEEE, ACS(CP))}
       
    \vspace{0.9cm}
    \textbf{Executive Summary}
\end{center}

AI is transforming the existing technology landscape at a rapid phase enabling data-informed decision making and autonomous decision making. Unlike any other technology, because of the decision-making ability of AI, ethics and governance became a key concern. There are many emerging AI risks for humanity, such as autonomous weapons,
automation-spurred job loss, socio-economic inequality, bias caused by data and algorithms,
privacy violations and \textit{deepfakes}. Social diversity, equity and inclusion are considered key success factors of AI to mitigate risks, create values and drive social justice. Sustainability became a broad and complex topic entangled with AI. Many organizations (government, corporate, not-for-profits, charities and NGOs) have diversified strategies driving AI for business optimization and social-and-environmental justice. Partnerships and collaborations become important more than ever for equity and inclusion of diversified and distributed people, data and capabilities. Therefore, in our journey towards an AI-enabled sustainable future, we need to address AI ethics and governance as a priority. These AI ethics and governance should be underpinned by human ethics.

\vspace{2pc}
\noindent{\it Keywords}: AI, Governance, Ethics, Sustainability, ESG (Environmental, Social, and Governance), SDGs (Sustainable Development Goals), DEI (Diversity, Equity, and Inclusion), Social Justice, Framework

\tableofcontents{}

\listoffigures

\let\cleardoublepage=\clearpage

\mainmatter

\pagebreak
\chapter{Introduction}

AI has been identified as the new electricity \cite{stanfordAndrewElectricity}. Data has been considered the oil for the digital economy. This is also considered the 4$^{th}$ industrial revolution. From this perspective, have we thought about the sustainability of the new electricity: AI?

When the steam engine was deployed in the 1$^{st}$ industry revolution and electricity was generated in the 2$^{nd}$ industrial revolution, sustainability had not been a concern. Humans' rush to economic advantages from the 1$^{st}$ and 2$^{nd}$ industrial revolutions caused many problems in the long run, such as climate change and related environmental and humanitarian crises \cite{Abram2016}. By the time we retrospect and think about the sustainability of power and energy generation, it has caused significant damage to humanity. Therefore, we mustn't be making the same mistake in the 4$^{th}$ industrial revolution: AI.

AI governance is a complex process as AI has autonomous decision-making capability. Consequently, AI can create fundamental risks in human dignity, human rights and human autonomy \cite{ArtoLaitinen2021}, \cite{Zardiashvili2020}, \cite{Boni2021}. Hence, AI ethics and governance must be realized from the very beginning when humans initiate artificial intelligence. Therefore AI ethics should be underpinned by human ethics \cite{mgsAfr2022}.

\chapter{Human Ethics}

\textit{Consequentialism} and \textit{Utilitarianism} can be identified as two broad categories of human ethics. Consequentialism is a theory that says whether something is ethical or not depends on its outcomes or consequences. In this way, the focus is on outcomes rather than the overall benefit or process.  In contrast, in Utilitarianism, the ethical nature is decided based on whether the process is optimised to maximise the overall benefit to the society rather than the outcomes. These two different ethical perspectives sometimes create a dilemma, where we may see a decision is ethical in the Consequentialism perspective but not ethical in the Utilitarianism perspective and vice versa. Therefore, the leaders need to understand both perspectives and make sure AI realisation can be justifiable in both perspectives as much as possible.  

\begin{figure*}[!htbp]
  \centering
  \captionsetup{justification=centering}
  \captionsetup{labelfont=bf}
  \includegraphics[width=0.8\textwidth]{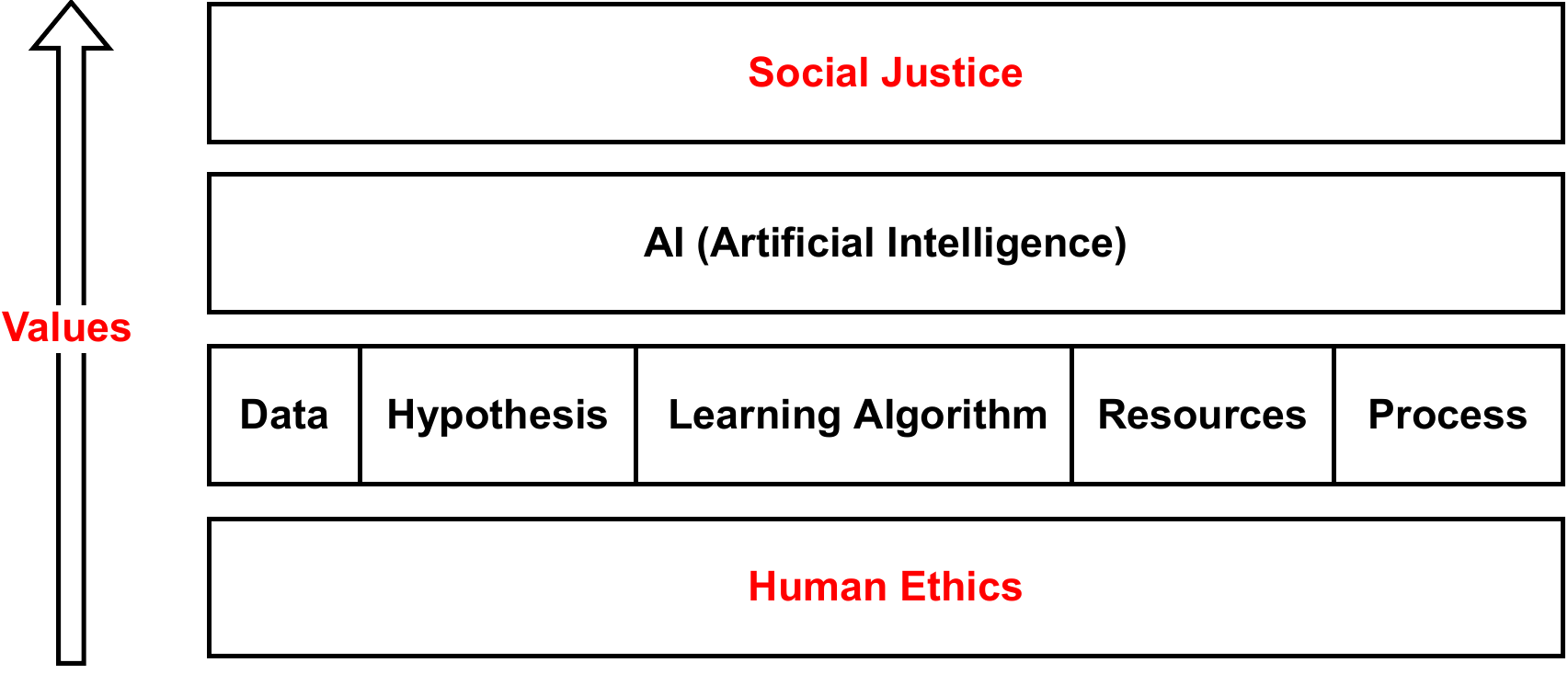}
  \caption{AI is a capability which can transform values of human, data and technologies towards social justice.}
  \label{fig:AI_Ethics_Intro}
\end{figure*}

Human should consider AI as a capability rather than an agent. AI should not take autonomy wherever human dignity is a concern. The fundamental purpose of AI is to transform the values of human, data and technologies towards social justice (see Figure~\ref{fig:AI_Ethics_Intro}) by optimising the Consequentialism and Utilitarianism perspectives of human ethics.

\begin{figure*}[!htbp]
  \centering
  \captionsetup{justification=centering}
  \captionsetup{labelfont=bf}
  \includegraphics[width=0.68\textwidth]{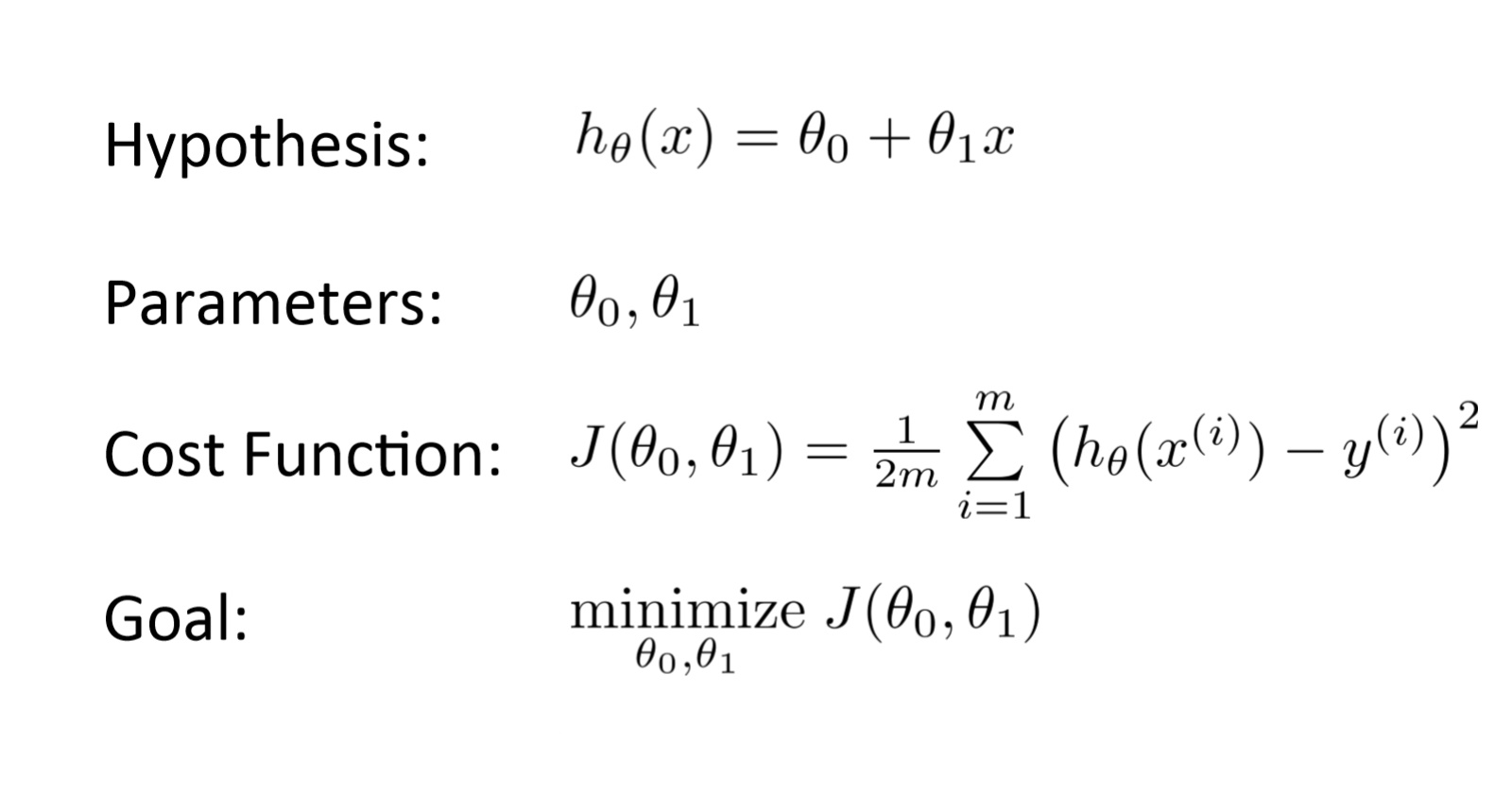}
  \caption{The algorithm of linear regression which fit a straight line cross the data points. Note that human decisions on selection of optimisation problem, data, algorithm, and parameters. How can we ethically govern these decisions?}
  \label{fig:AI_linear_regression}
\end{figure*}

In technical perspective, humans are accountable for their decisions on AI implementations:
\begin{itemize}
    \item bias mitigation, 
    \item problem selection, 
    \item opportunity cost evaluation for social justice, 
    \item data selection and sampling, 
    \item insight (features) incorporation, 
    \item algorithm selection, 
    \item hyperparameter tuning,
    \item regularisation, etc.
\end{itemize}
Figure~\ref{fig:AI_linear_regression} shows the basic touch-points of human decision making in a simple form of an AI algorithm, \textit{linear regression}. Note how human decision-making influences a typical AI solution in data, hypothesis, algorithmic, resource and process perspectives. Many tools (e.g., MLOps, ModelOps, AIOps, XOps, DataOps) enable and facilitate deciding and fine-tuning all of those factors and aspects. Our ethics, knowledge and risk appetite determine \textbf{why how and what} we do, which is why AI governance and ethics are important.

\chapter{AI from the Consequentialism Perspective}

\begin{figure*}[!htbp]
  \centering
  \captionsetup{justification=centering}
  \captionsetup{labelfont=bf}
  \includegraphics[width=\textwidth]{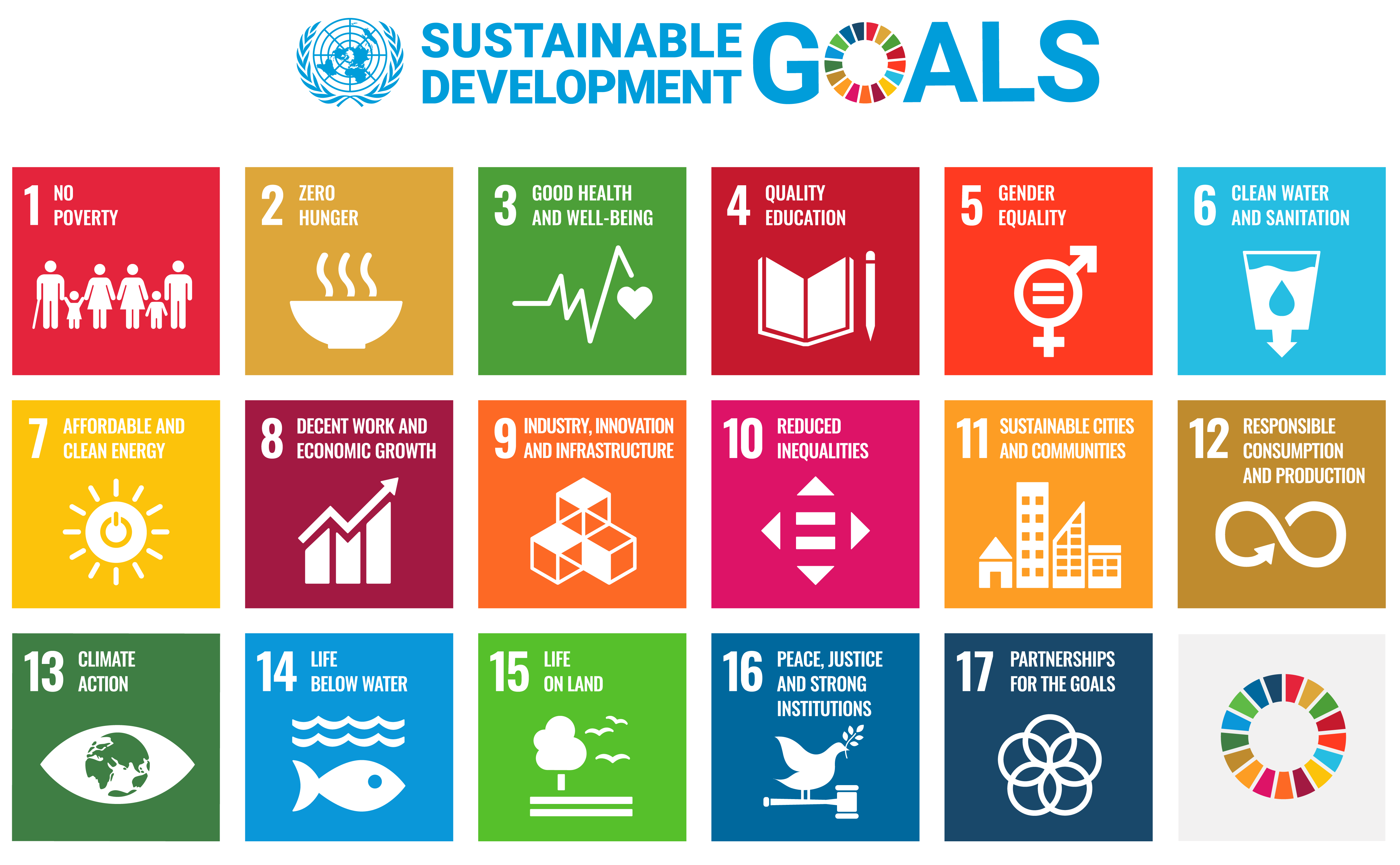}
  \caption{UN Sustainable Development Goals (SDGs) \cite{cf2015transforming}. In 2015, United Nations member states adopted these 17 SDGs as their 2030 agenda for sustainable development.}
  \label{fig:UN_SDGs}
\end{figure*}

AI can support 79\% of the United Nations 17 Sustainable Development Goals (SDGs) (see Figure~\ref{fig:UN_SDGs}) \cite{Vinuesa2020}, which is the foundation of ESG and Social Impact strategies planned to realise by 2030. In 2015, United Nations member states adopted these 17 SDGs as their 2030 agenda for sustainable development \cite{cf2015transforming}. This agenda establishes a shared framework for peace and prosperity for a sustainable future for people and the planet. The framework supports environmental, social and corporate governance (ESG) for sustainability.

\begin{figure*}[!htbp]
  \centering
  \captionsetup{justification=centering}
  \captionsetup{labelfont=bf}
  \includegraphics[width=0.9\textwidth]{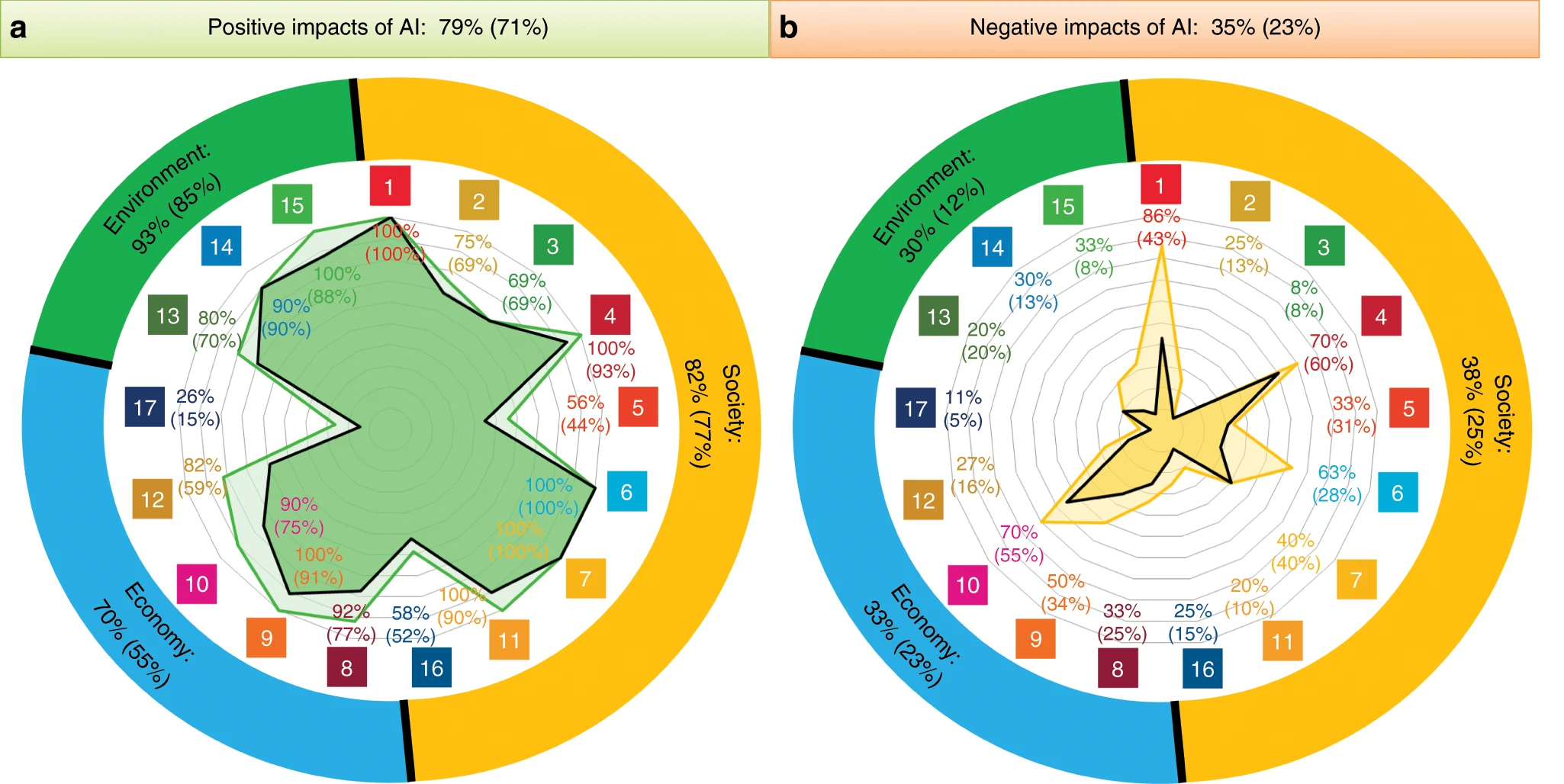}
  \caption{Analysis of positive and negative impact of AI on the UN SDGs \cite{Vinuesa2020}. Figure courtesy of \cite{Vinuesa2020}.}
  \label{fig:AI_for_SDGs}
\end{figure*}

In \textit{Consequentialism} perspective of AI ethics, UN SDGs provide a globally acceptable ethical framework for AI governance. However, depending on governance and ethics of AI, there can be pros and cons in AI applications. Figure~\ref{fig:AI_for_SDGs} shows how AI impacts positively and negatively on each UN SDGs.

\begin{figure*}[!htbp]
  \centering
  \captionsetup{justification=centering}
  \captionsetup{labelfont=bf}
  \includegraphics[width=0.9\textwidth]{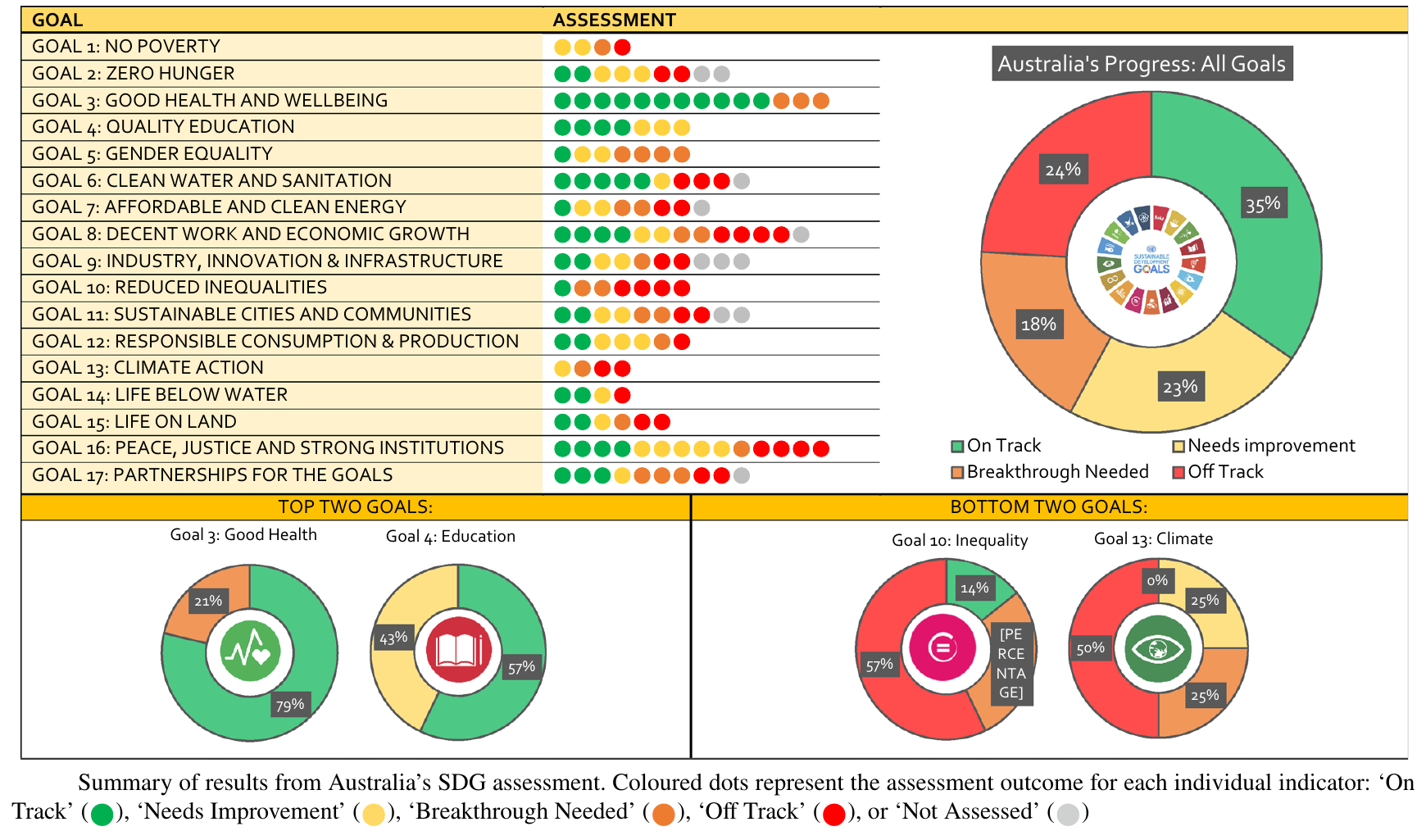}
  \caption{Results of Australia’s SDG assessment \cite{Allen2019}. Note the goals in which Australia is \textit{off track} and needs \textit{breakthrough}. Figure courtesy of \cite{Allen2019}.}
  \label{fig:AUS_SDGs_performance}
\end{figure*}

\begin{figure*}[!htbp]
  \centering
  \captionsetup{justification=centering}
  \captionsetup{labelfont=bf}
  \includegraphics[width=\textwidth]{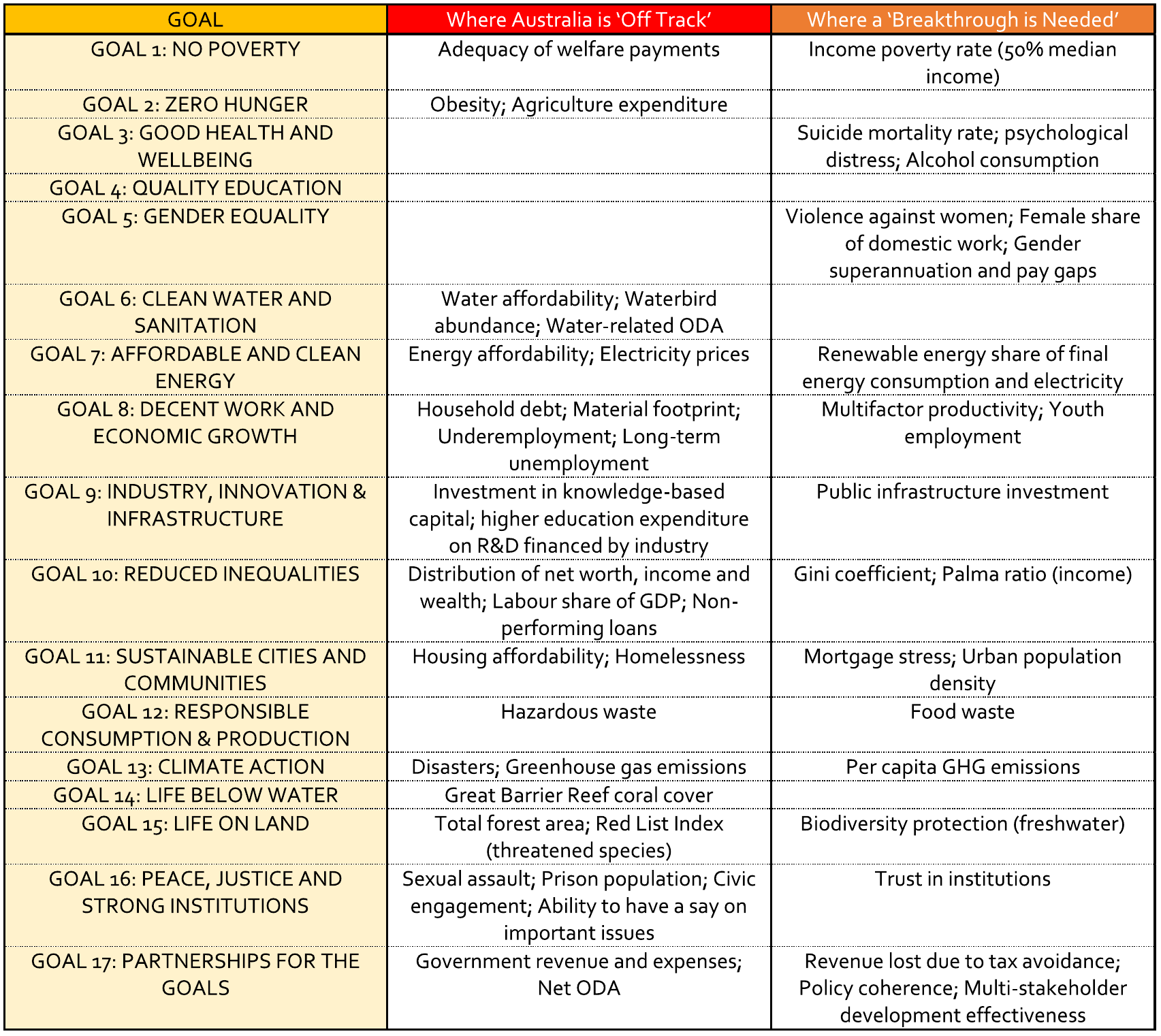}
  \caption{Risk landscape of Australia’s SDGs \cite{Allen2019}. Australia need to focus these concerns aligning with accelerated economic developments. Figure courtesy of \cite{Allen2019}.}
  \label{fig:AUS_SDGs_Risks}
\end{figure*}

The UN SDGs are an urgent call for action by all countries - developed and developing - in a global partnership; the Australian organisations must address this diligently. Australia still has a long journey ahead in achieving UN SDGs. Figure~\ref{fig:AUS_SDGs_performance} illustrates the results of Australia’s SDG assessment \cite{Allen2019}. Note the goals in which Australia is \textit{off track} and needs a \textit{breakthrough}. 
Moreover, Figure~\ref{fig:AUS_SDGs_Risks} summarises the Australian concerns related to unsatisfactory progress in each UN SDG analysed in \cite{Allen2019}.
Therefore, in the economic acceleration effort with AI, the government should focus on achieving UN SDGs effectively, which will promote AI ethics, governance and \textit{AI for sustainability}.

\chapter{AI from the Utilitarianism Perspective}

In the Utilitarianism perspective of AI ethics and governance, the motivation would be to maximise the overall benefit to the society instead of morality. In this perspective, leaders are encouraged to look into the more granular level and customised design and implementations rather than premeditated norms, moral conventions or solutions (which are more focused on the Consequentialism perspective). The following are important design concerns when focusing on AI ethics and \textit{sustainability of AI} from the Utilitarianism perspective. 

\section{Bias}

\begin{figure*}[!htbp]
  \centering
  \captionsetup{justification=centering}
  \captionsetup{labelfont=bf}
  \includegraphics[width=0.8\textwidth]{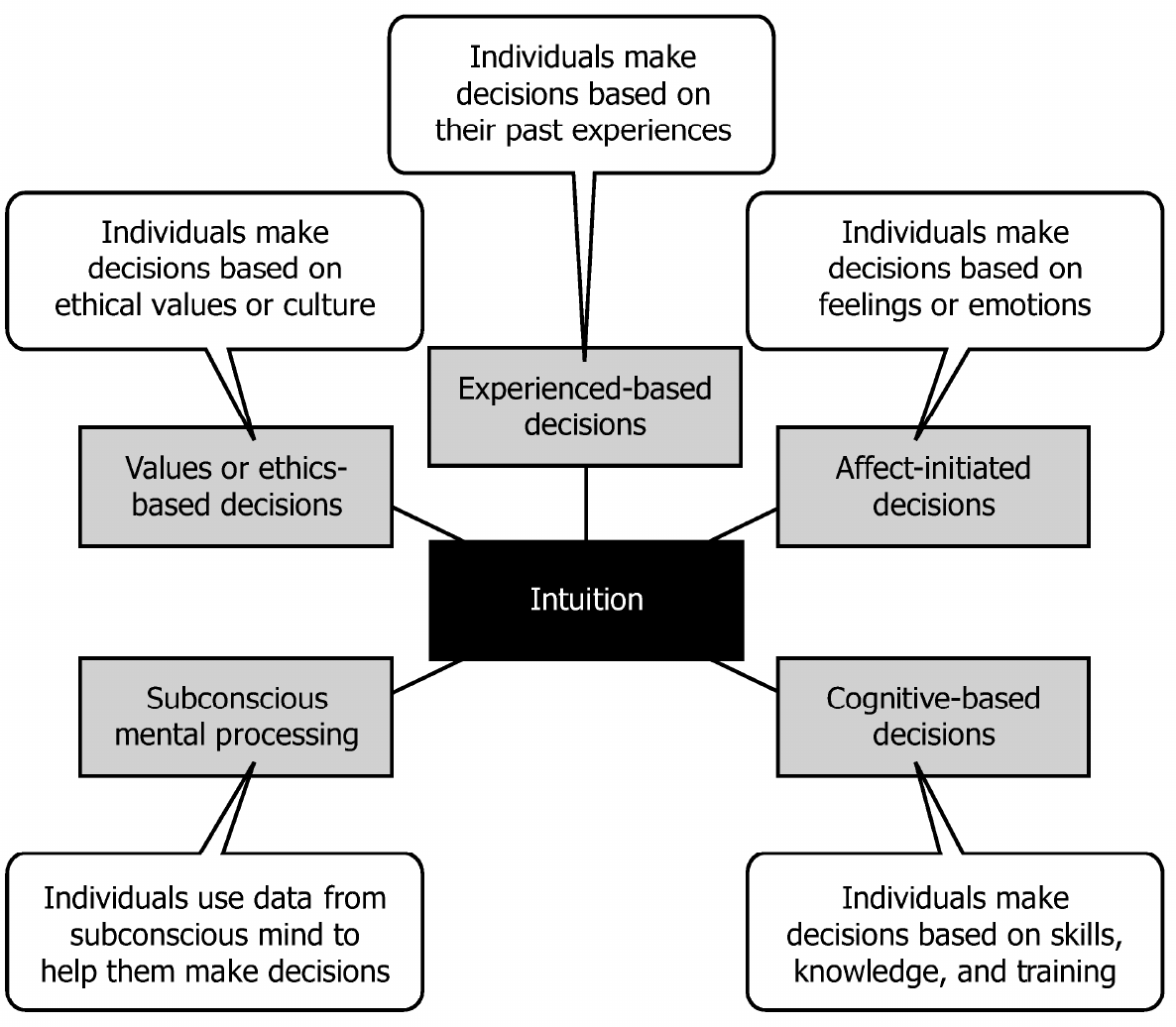}
  \caption{The nature of intuitive decision-making \cite{Burke1999}. Figure \textcopyright Australian Institute of Company Directors.}
  \label{fig:intuition}
\end{figure*}

Bias in data, algorithms and people is the fundamental cause of the failure of AI implementations. Unlike many other applications, AI is introduced to involve autonomous, semi-autonomous or prescriptive decision making. Therefore, it is important to mitigate the biases in AI to maximise social justice. The leaders should be self-aware, conscious, and avoid intuitive decisions on AI implementations, management and governance. Figure~\ref{fig:intuition} shows the traits of intuitive decision-making. The collaborations, partnerships and working as a distributed network are recommended by the 17$^{th}$ UN SDGs to overcome those traits by promoting diversity, equity and inclusion in people realising AI.

\begin{figure*}[!htbp]
  \centering
  \captionsetup{justification=centering}
  \captionsetup{labelfont=bf}
  \includegraphics[width=0.75\textwidth]{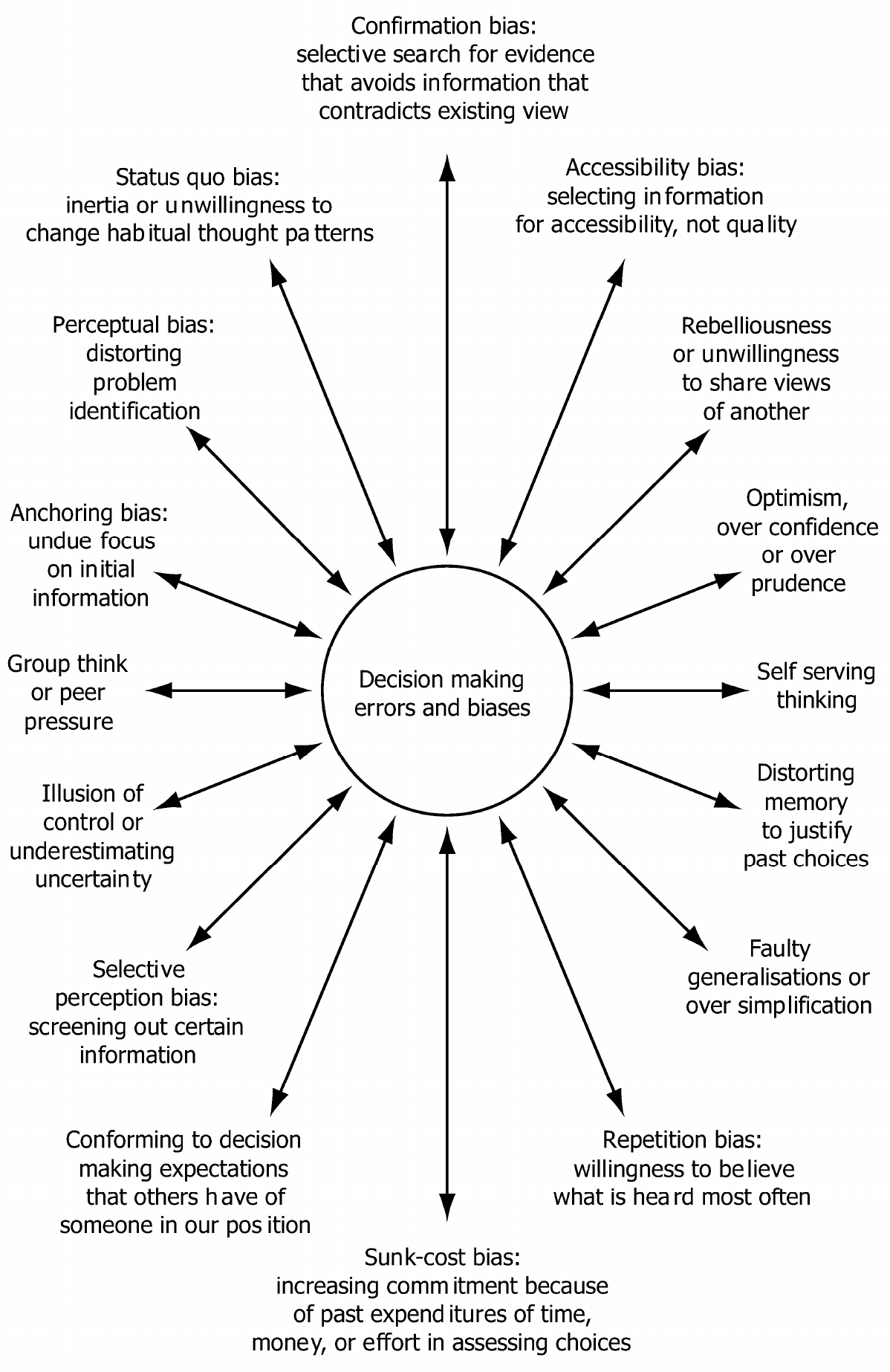}
  \caption{Decision making errors and biases \cite{hammond1998hidden}. Figure \textcopyright Australian Institute of Company Directors.}
  \label{fig:biases}
\end{figure*}

It is understood that each individual has their own biases, traits and ways of thinking. That is why collective decision making with a diverse group is more effective than individual decision making. Figure~\ref{fig:biases} shows various decision-making errors and biases that leaders should be aware of when forming, norming and driving AI strategies and transformation. Diverse perspectives, more information, more alternatives, and different thinking styles are key success factors of Utilitarianism perspectives of AI ethics, which help democratise AI, avoiding disparities and meaningful participation and representation \cite{Stahl2021}.

\section{Diversity}

\begin{figure*}[!htbp]
  \centering
  \captionsetup{justification=centering}
  \captionsetup{labelfont=bf}
  \includegraphics[width=\textwidth]{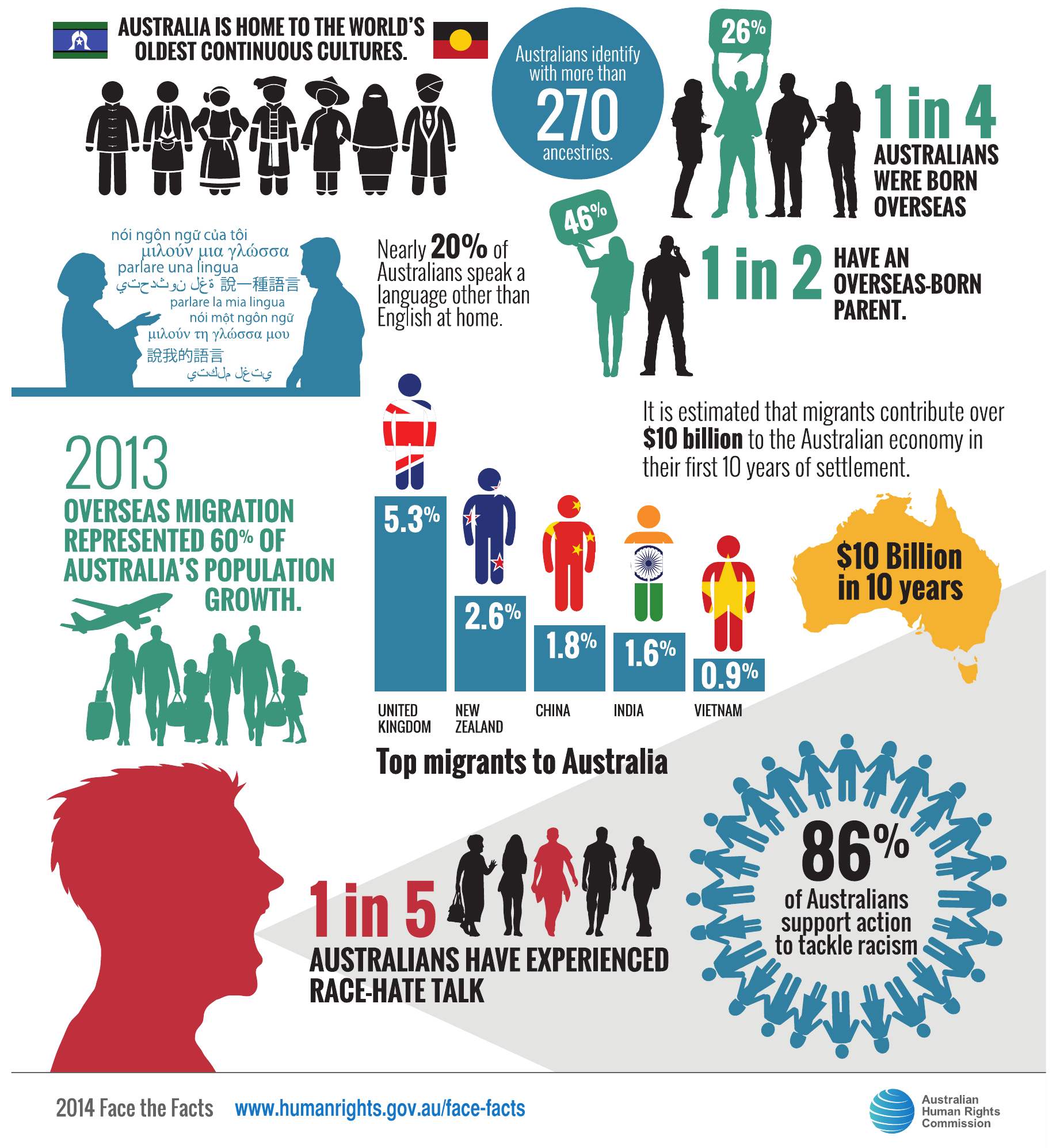}
  \caption{Cultural diversity of Australia and interesting facts. Figure \textcopyright Australian Human Rights Commission \cite{AHRC-online}.}
  \label{fig:AUS_Cultural_Diversity}
\end{figure*}

Australia has vibrant multicultural community (see Figure~\ref{fig:AUS_Cultural_Diversity}). This is one of the uniqueness of Australia. The Aboriginal and Torres Strait Islander peoples’ culture is the world’s oldest continuous culture. Australians can be related to more than 270 ancestries. Since 1945, almost 7 million people have migrated to Australia. This rich culture is one of the greatest strengths of its economic prosperity. Therefore, it is important to consider this great diversity when mitigating biases and promoting inclusions in AI initiatives. 

Leaders should bring diversity to AI solutions by enabling equity and inclusion. {\normalsize \textit{“Neither a person nor an apple can be diverse. Diversity is the property of a collection of people—a basket with many kinds of fruit”}}\cite{SEPage2007}. Gender equality and reduced inequalities are key focuses in sustainability addressing through 5$^{th}$ and 10$^{th}$ UN SDGs. On the other hand, the Australian anti-discrimination law was established to eliminate all forms of discrimination which is an integral part of promoting diversity \cite{au-anti-discrimination}.

\section{Impartiality and Localisation}

Impartiality and localisation are two important objectives in an equitable AI solution. When managing impartiality, retaining fairness to locality is equally important. If the AI model is generalised across the entire population, it may be justified as an impartial solution but might not be fair for minority groups. Even deploying locally optimised multiple models may create injustice to people at the margins of the segments and cause issues from the impartiality perspective. 

\begin{figure*}[!htbp]
  \centering
  \captionsetup{justification=centering}
  \captionsetup{labelfont=bf}
  \includegraphics[width=0.75\textwidth]{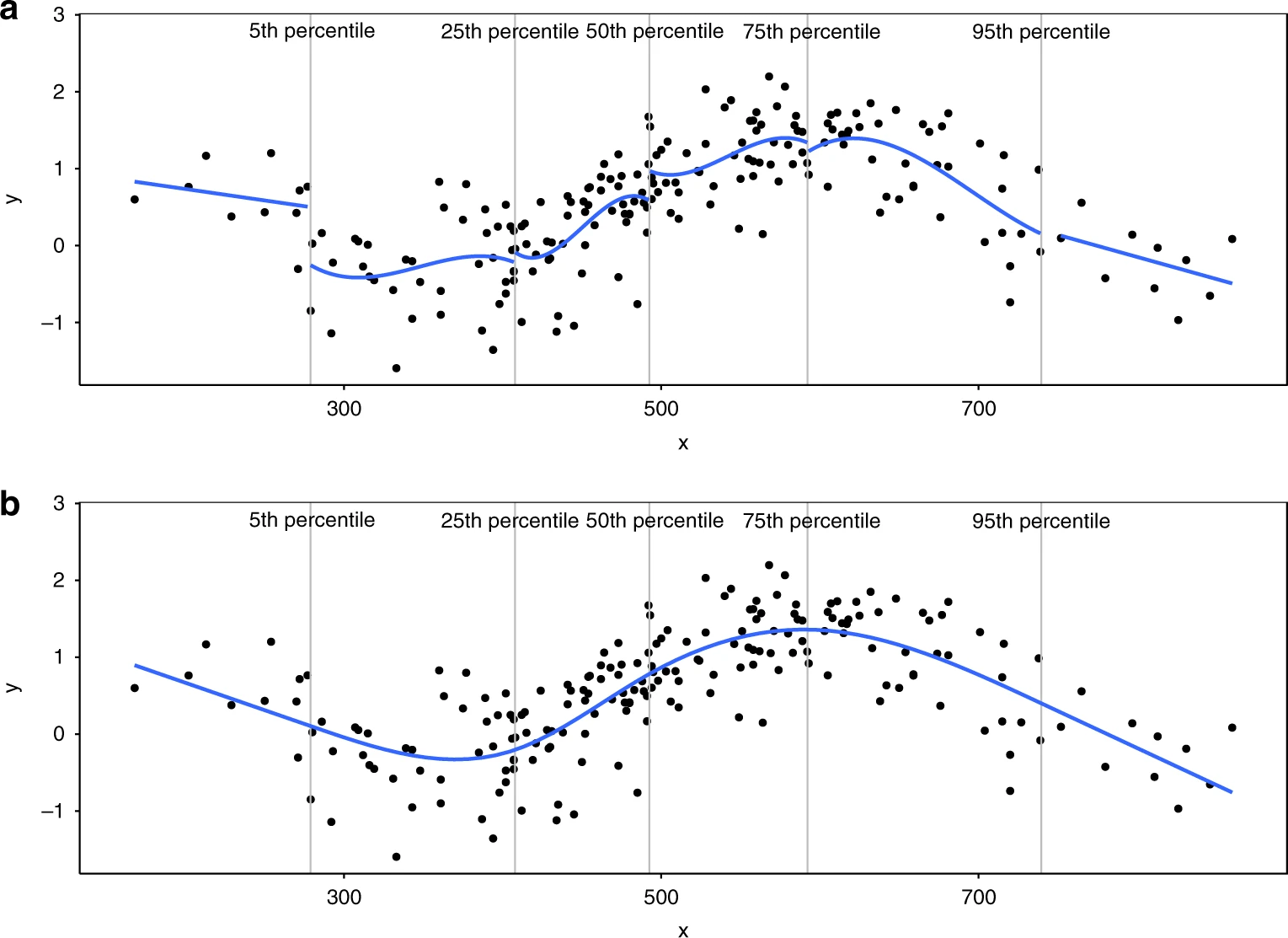}
  \caption{Categorising continuous variables is important for diversifying the service. Modelling of each category of continuous variable independently, as shown in the figure \textbf{a} can lead to loss of information and poor predictions. On the other hand, modelling the entire data set with a single higher-order polynomial might overfit the model. The figure \textbf{b} shows mathematically complex restricted-cubic-spline regression lines, which can flexibly and accurately model complex and non-linear relationships \cite{Gauthier2019}.}
  \label{Fig:restricted_cubic_spline}
\end{figure*}

Figure~\ref{Fig:restricted_cubic_spline} shows two modelling strategies on complex and diversified data points. In machine learning, regularisation techniques generalise the model while mitigating overfit. Sometimes, the regularisation may neglect the minority requirements. Therefore, the model complexity on data should be determined by accounting impartiality and localisation of the solution.

\section{Equity}

\begin{figure*}[!htbp]
  \centering
  \captionsetup{justification=centering}
  \captionsetup{labelfont=bf}
  \includegraphics[width=0.9\textwidth]{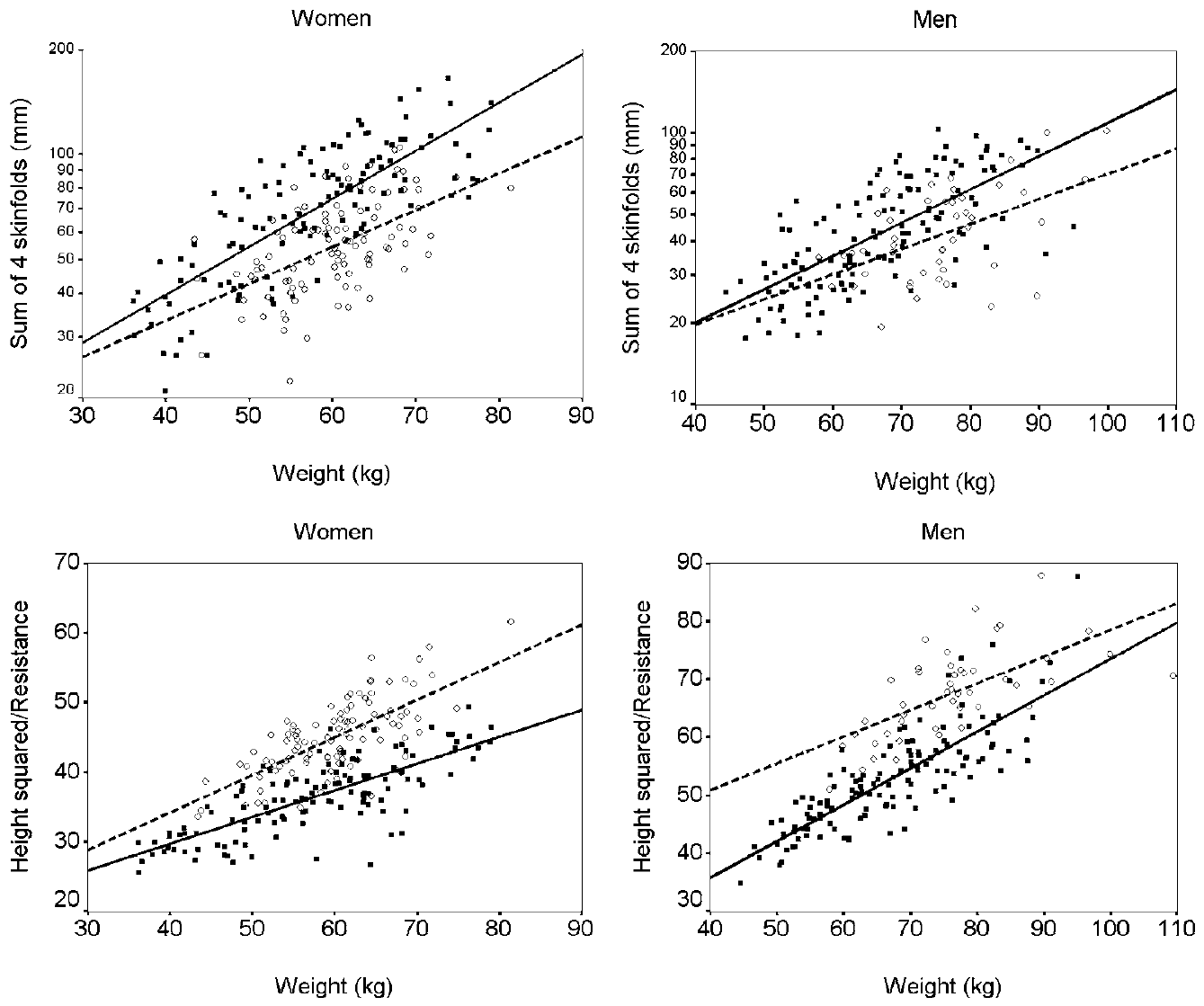}
  \caption{Relationship of body weight to surrogate measures of fat mass (sum of four SFT) and fat-free mass (height$^2$/resistance) in Australians of Aboriginal (filled squares, solid line) and European (open circles, broken line) ancestry \cite{PMID:12879090}.}
  \label{Fig:BMI_FNP}
\end{figure*}

Equity is an important concern in social justice, which is quite relevant to the Australian multicultural society. Bringing AI equity to relevant groups is important when creating values or making decisions from an ethical perspective. For example, Aboriginal and European Australians have a significantly different body fat distribution and fat mass for given body weight or BMI. By research, it has been identified that (see Figure~\ref{Fig:BMI_FNP}) BMI ranges valid for the majority of Australians to determine weight status may be inappropriate in Australian Aboriginal people \cite{PMID:12879090}.

\section{Inclusion}

\begin{figure*}[!htbp]
  \centering
  \captionsetup{justification=centering}
  \captionsetup{labelfont=bf}
  \includegraphics[width=\textwidth]{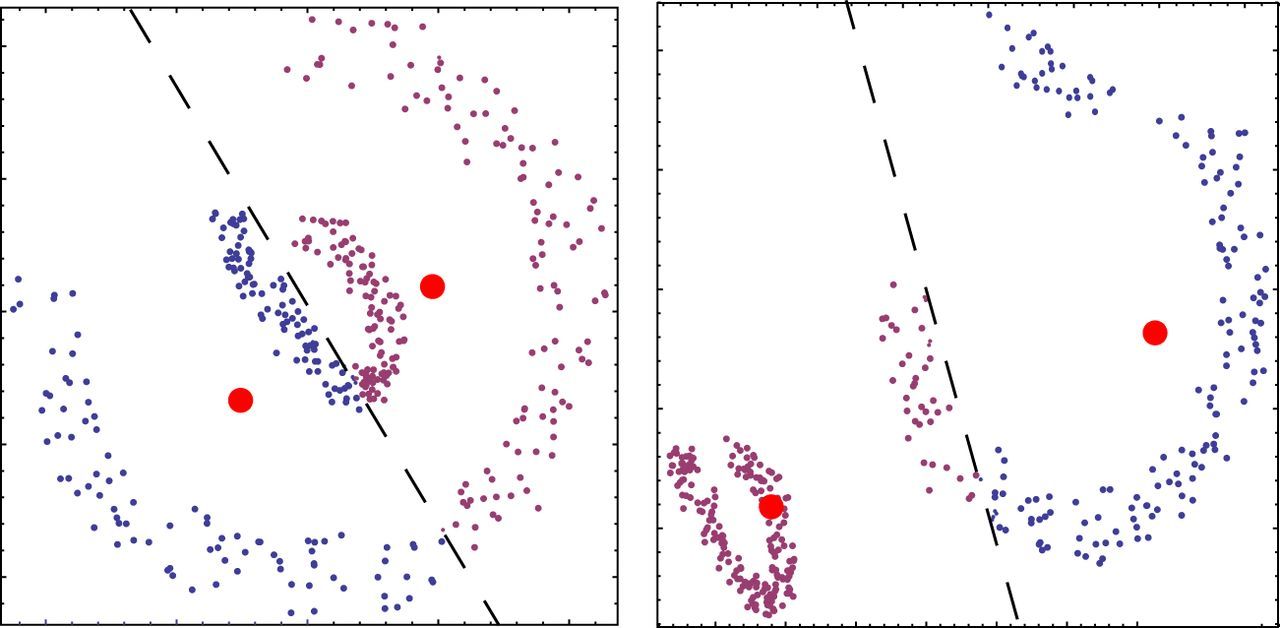}
  \caption{How K-means clustering getting unsuccessful in non-Gaussian data distribution. The dashed line denotes separating the computed cluster boundaries; filled dots, cluster centres \cite{Lorimer2017ClusteringHM}. By bringing reasonable insight, the K-means clustering can be enhanced.}
  \label{Fig:Bias_Clustering}
\end{figure*}

Reducing \textit{overfit} of an AI algorithm by regularisation and/or dimensionality reduction may disregard important attributes related to minority groups. Therefore, data scientists should bring the right amount of data insights to the design to enhance inclusiveness, which can be considered a controlled bias. For example, most of the time, the initiation of hyperparameters is important at the start of unsupervised learning. This intentional bias can enhance the quality of an AI solution. Poor control of machine learning is difficult to be compensated for and can lead to undesirable outcomes (see Figure~\ref{Fig:Bias_Clustering}) \cite{Lorimer2017ClusteringHM}. 

\chapter{Complexity in AI Governance}

The AI spectrum is quite broad \cite{mgsGaiei2022}. From IoT sensor management to smart city development, different stakeholders should look into different perspectives such as social justice, strategy, technology, sustainability, ethics, policies, regulations, compliance, etc. Moreover, things get even more complex when different perspectives are entangled. As examples,
\begin{enumerate}
    \item Environmental and Social: AI has been identified as a key enabler on 79\% (134 targets) of United Nations (UN) Sustainable Development Goals (SDGs) \cite{Vinuesa2020}. However, 35\% (59 targets) may experience a negative impact from AI. While the environment gets the highest potential, the society gets the most negative impact from AI and creates social concerns,
    \item Environmental and Technology: Cloud computing is promising with the availability and scalability of resources in data centres. With emerging telecommunication technologies (e.g., 5G), the energy consumption when transferring data from IoT/edge devices to the cloud became a concern on carbon footprint and sustainability. This energy concern is a factor that shifts the technology landscape from cloud computing to fog computing \cite{Baccarelli2017},
    \item Economic and Sustainability: Businesses are driving AI, hoping it can contribute about 15.7 trillion to the world economy by 2030 \cite{PwC_AI_2017}. On the other hand, the UN SDGs are also planned to achieve by 2030 in the areas critically important for humanity, and the planet \cite{cf2015transforming}. The synergy between AI economic and sustainability strategies will be essential,
    \item Economic and Social: Businesses are driving AI, hoping it can contribute about 15.7 trillion to the world economy by 2030. However, the research found that 85\% of AI projects will fail due to bias in data, algorithms, or the teams responsible for managing them \cite{Gartner2018}. Therefore, AI ethics and governance for the sustainability of AI became a key success factor in economic goals in AI.
    \item Economic and Ethical: Still, no government has been able to pass AI law except ethical frameworks or regulatory guidelines \cite{Floridi2019}. Therefore, there are many emerging AI risks for humanity on our way to economic prosperity, such as autonomous weapons, automation-spurred job loss, socioeconomic inequality, bias caused by data and algorithms, privacy violations, and deepfakes \cite{Perc2019}.
\end{enumerate}
On the other hand, the complex differences in AI applications don’t necessarily mean there are no similarities in other perspectives such as cultural values, community or strategy. For example, similar organizations may work on different sustainability goals for social justice. Such differences in AI strategy should not obstruct the partnership and collaboration opportunities between them.

\chapter{A Framework and a Model for AI Governance}

When addressing AI governance requirements, the complexity of the AI can be identified as the main challenge \cite{mgsGaiei2022}. Unlike any other technology, AI governance is complex because of its autonomous decision-making capability and influence on people's decision-making. Hence, AI governance is entangled with human ethics, which must be realised where artificial intelligence is applied or influenced. We introduced a framework and model with the simple golden circle in mind. They help directors find solutions for why, how and what questions when governing AI. First, the innovative KITE conceptualised abstraction framework helps directors drive the purpose of AI initiatives to address key success factors. With the support of the KITE abstraction framework, the innovative Wind-turbine conceptualised model helps to develop a comprehensive AI strategy for organisations. These frameworks and models help drive AI for sustainability in more structured, systematic, transparent, and collaborative ways.

\section{KITE abstraction framework}

\begin{figure*}[!htbp]
  \centering
  \captionsetup{justification=centering}
  \captionsetup{labelfont=bf}
  \includegraphics[width=0.8\textwidth]{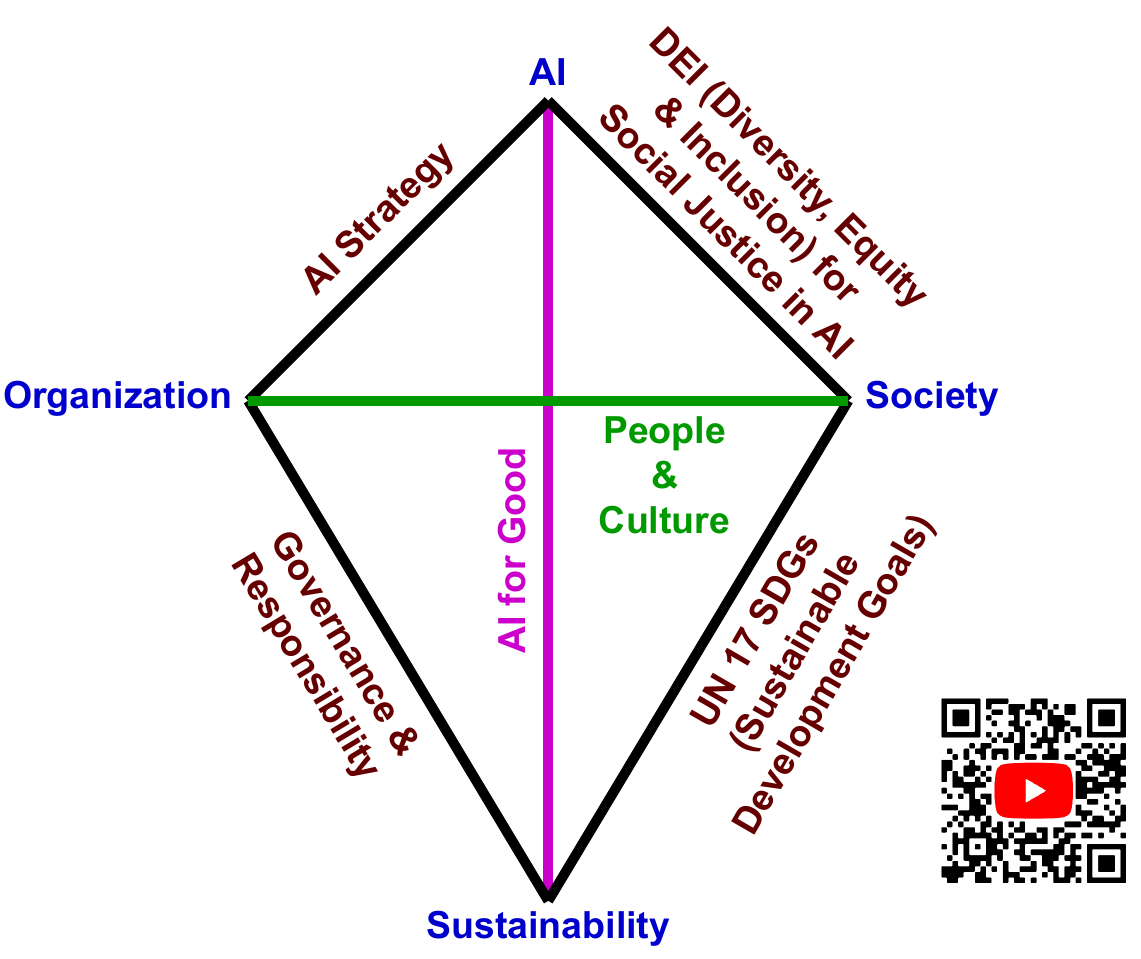}
  \caption{KITE abstraction framework for AI governance \cite{mgsUn2021}. It aligns with the broader ESG purpose, fundamentally the \textit{why} aspect of the golden circle.}
  \label{fig:KITE}
\end{figure*}

The KITE abstraction framework (see Figure~\ref{fig:KITE}) \cite{mgsUn2021} helps directors govern AI aligning with the broader ESG purpose, fundamentally the \textit{\textbf{why}} aspect of the golden circle. Irrespective of the complexity of the AI application, this framework analyses the four key dimensions of
\begin{enumerate}
    \item[1.] AI,
    \item[2.] Organisation,
    \item[3.] Society, and 
    \item[4.] Sustainability.
\end{enumerate}
The interdependencies of these dimensions enable addressing of AI strategy, AI for Good and United Nations Sustainable Development Goals. Further, it helps mitigate AI risks due to biases by bringing social diversity, equity and inclusion to AI governance. As illustrated in the diagram, it helps organisational governance and responsibilities by guiding the orchestration of people, culture and AI mission towards sustainability.

\pagebreak

\section{Wind-turbine conceptualised model}

\begin{figure*}[!htbp]
  \centering
  \captionsetup{justification=centering}
  \captionsetup{labelfont=bf}
  \includegraphics[width=0.82\textwidth]{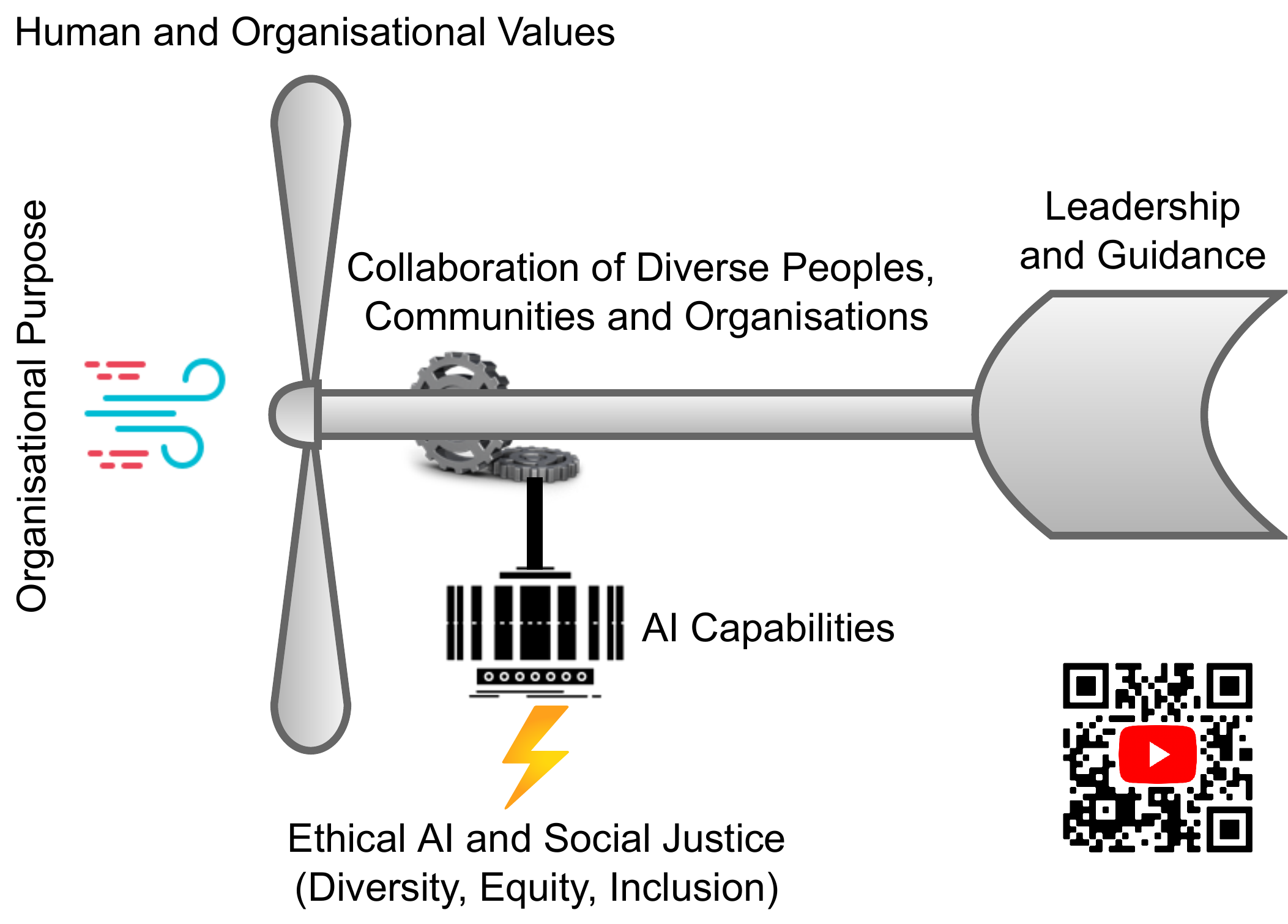}
  \caption{Wind-turbine conceptualised model for AI governance \cite{mgsHyper2021}. It helps directors address \textit{how} and \textit{what} aspects of AI governance.}
  \label{fig:Wind-turbine}
\end{figure*}

The wind-turbine conceptualised model (see Figure~\ref{fig:Wind-turbine}) \cite{mgsHyper2021} helps directors address \textit{\textbf{how}} and \textit{\textbf{what}} aspects of AI governance. The model helps oversee AI processes supporting social justice with social diversity, equity and inclusion. From the organisational perspective, this model directs the AI initiative towards humanity and sustainable development goals (SDGs) for minimising human suffering. Further, this model helps oversee the operations and management, represented by the tail of the wind turbine. The front-faced multi-blade rotor represents the values and policies (e.g., seven fundamental principles) that ethically and efficiently address humanitarian needs, risks and suffering. The wheels in the gearbox represent the community, partners and volunteers who are continually helping with diversity, equity and inclusion. Finally, the generator represents the Data and AI capabilities that drive the AI innovation and transformation for sustainability. In summary, directors can oversee the full spectrum of the AI processes, stakeholders, and management.

\section{People, Culture and Mission}

To make sure AI for good programs serve the purpose of serving humanity and sustainability, it is important to mitigate the biases in decision making in leadership, management and governance while managing the projects that enhance social justice. These make sure we can realise AI ethics and sustainable development goals.

However, to minimise biases and enhance social justice, it is required to bring social diversity, equity and inclusion to the leadership, management and governance. Only then can we achieve utilitarianism and consequentialism perspectives of human ethics which can underpin the AI ethics for serving humanity and sustainability. Our framework helps all stakeholders including communities, volunteers and partners to collaborate on sustainable development goals and social justice. 

From the corporate governance and management perspective, this framework helps the corporate board, human resource (HR) and management to orchestrate culture, people and mission towards humanity and sustainability. Figure~\ref{fig:Wind-turbine} illustrates how the synergy between corporate culture, people and mission can drive AI ethics towards sustainable AI and goals \cite{mgsHyper2021}.
\begin{figure*}[!htbp]
  \centering
  \captionsetup{justification=centering}
  \captionsetup{labelfont=bf}
  \includegraphics[width=0.6\textwidth]{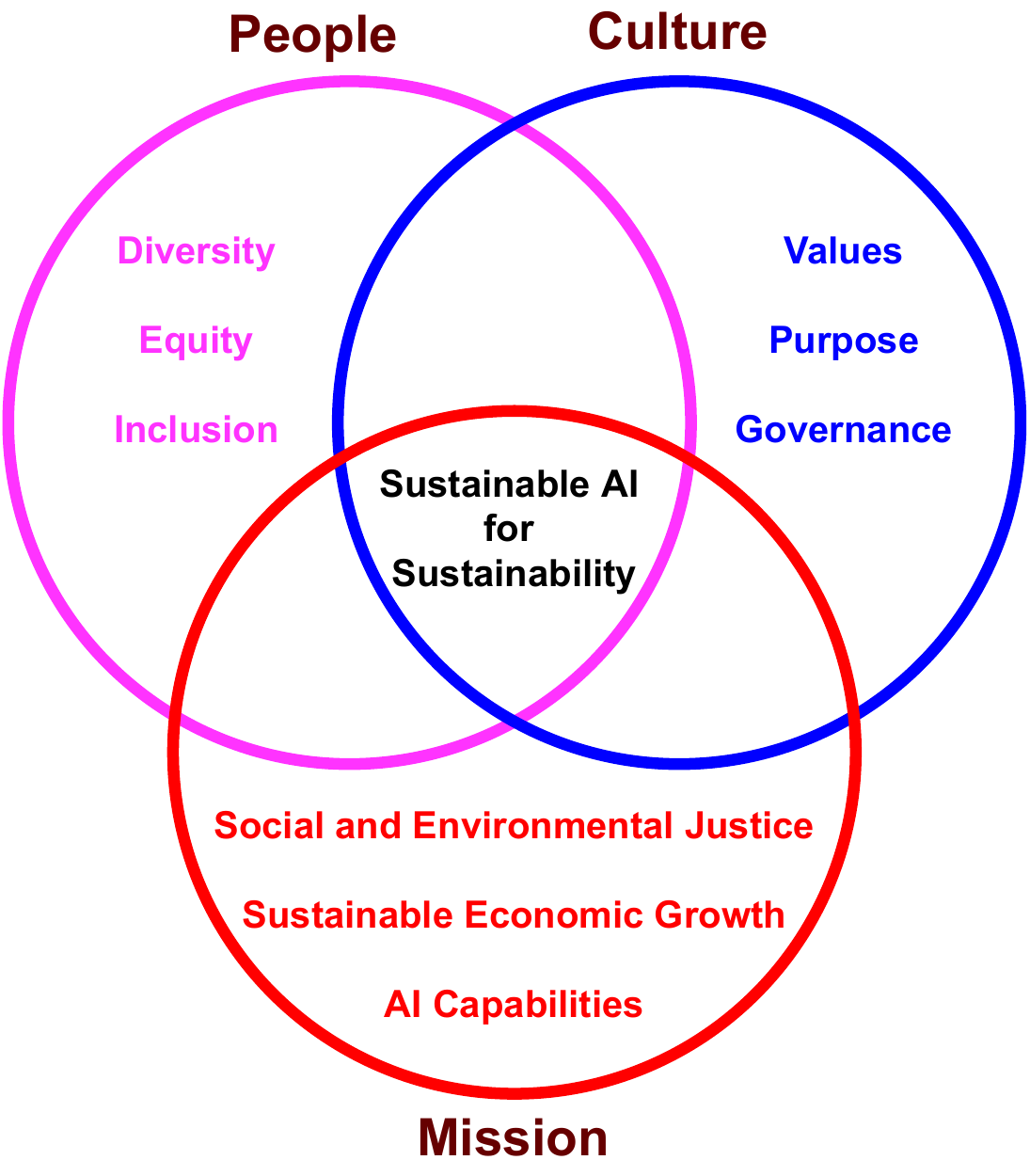}
  \caption{The proposed AI governance tools help the corporate board, human resource (HR) and management to orchestrate culture, people and mission towards humanity and sustainability.}
  \label{fig:AI_venn_diagram}
\end{figure*}

\chapter{Adaptation of the Framework}

\begin{figure*}[!htbp]
  \centering
  \captionsetup{justification=centering}
  \captionsetup{labelfont=bf}
  \includegraphics[width=\textwidth]{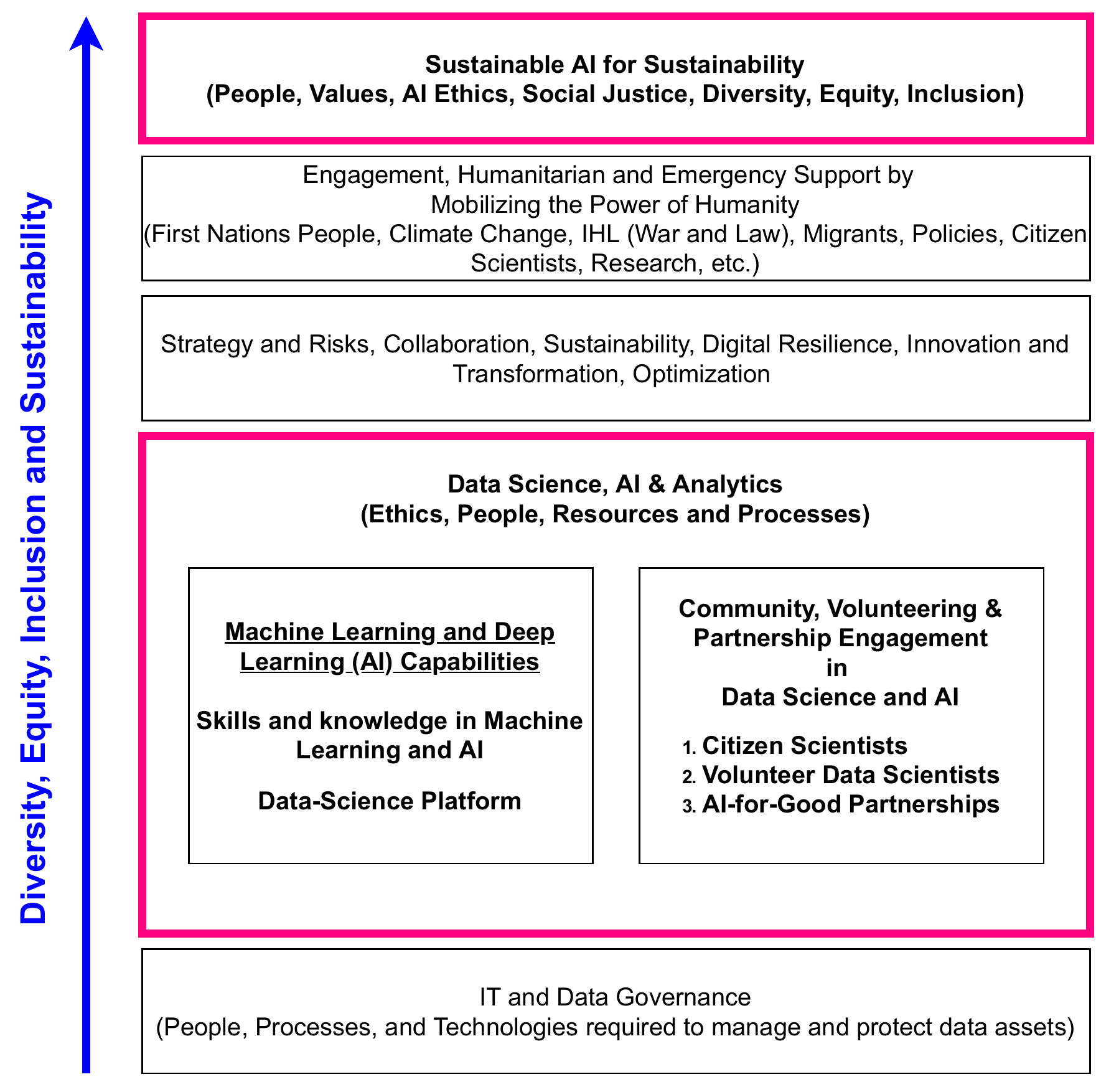}
  \caption{Sustainable AI for Sustainability. Businesses should position their IT, data science, and AI capabilities to address social justice and sustainability strategies. DEI (diversity, equity and inclusion) would be a key success factor of those initiatives.}
  \label{Fig:DEI_in_AI}
\end{figure*}

The adaptation of the proposed framework helps creating values based on the data wisdom \cite{mgsAicd2022}. It helps AI innovation and transformation towards social justice. As shown in Figure~\ref{Fig:DEI_in_AI} \cite{mgsHyper2021}, the data science and AI as a service layer supports business strategies of ESG by leveraging data and IT assets while enhancing DEI (diversity, equity and inclusion), brand advocacy, customer experience (CX), and return on investment (ROI).

\begin{figure*}[!htbp]
  \centering
  \captionsetup{justification=centering}
  \captionsetup{labelfont=bf}
  \includegraphics[width=\textwidth]{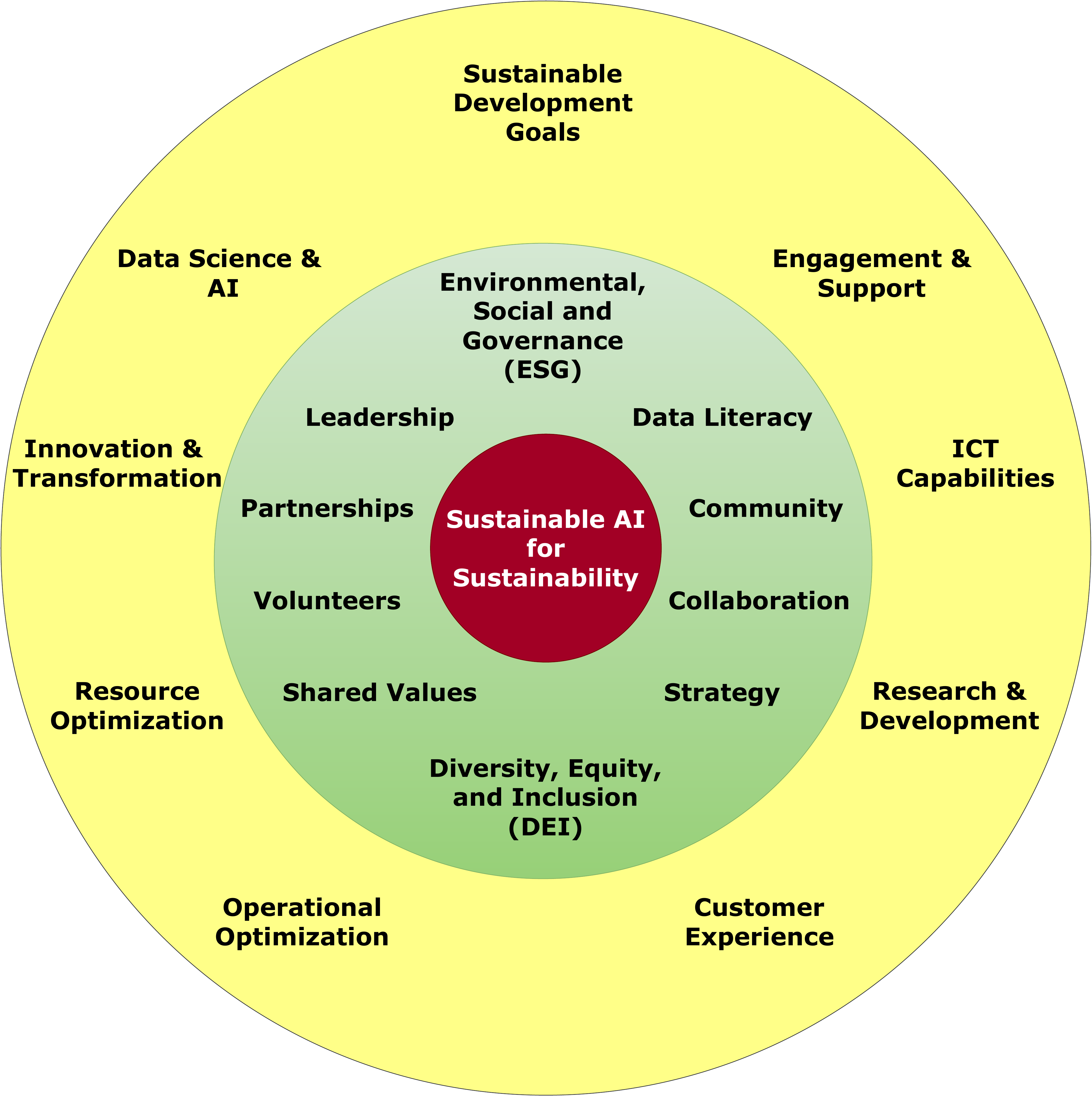}
  \caption{Development of sustainable AI as a core competency. AI has been identified as a key enabler for ESG and sustainability.}
  \label{Fig:The_AI_Golden_Circle}
\end{figure*}

The proposed framework establishes synergy between AI governance and social justice by mobilizing the organizational culture towards AI-driven innovation and transformation. A greater social diversity, equity and inclusion can be expected in AI initiatives which enable ethical inclusion, processes and outcomes in AI. The sustainable AI and sustainable development goals will be a primary focus in AI developments that drive business objectives and corporate social responsibilities. The ideation of this strategy can be illustrated by Figure~\ref{Fig:The_AI_Golden_Circle} \cite{mgsHyper2021}.

\chapter{Conclusion}

AI would be a key capability for future prosperity. Good governance of AI is very important to mitigate AI risks and create values. AI frameworks and standards are emerging to govern AI aligning with human ethics and emerging environmental, social, and corporate governance (ESG) principles. In brief, diversity, equity and inclusion (DEI) together with social and cultural values can make AI initiatives vibrant and sustainable. Further, it will mitigate biases related to AI, including biases in data, algorithms, people, and processes. This book's recommendations will help leaders orchestrate people, culture, and mission toward sustainable AI for social justice.

\pagebreak
\renewcommand{\bibname}{References}
\bibliographystyle{main}
\bibliography{main}

\pagebreak
\section*{Author's Biography}
\begin{wrapfigure}{l}{0.3\textwidth} 
\includegraphics[width=0.28\textwidth,clip,keepaspectratio]{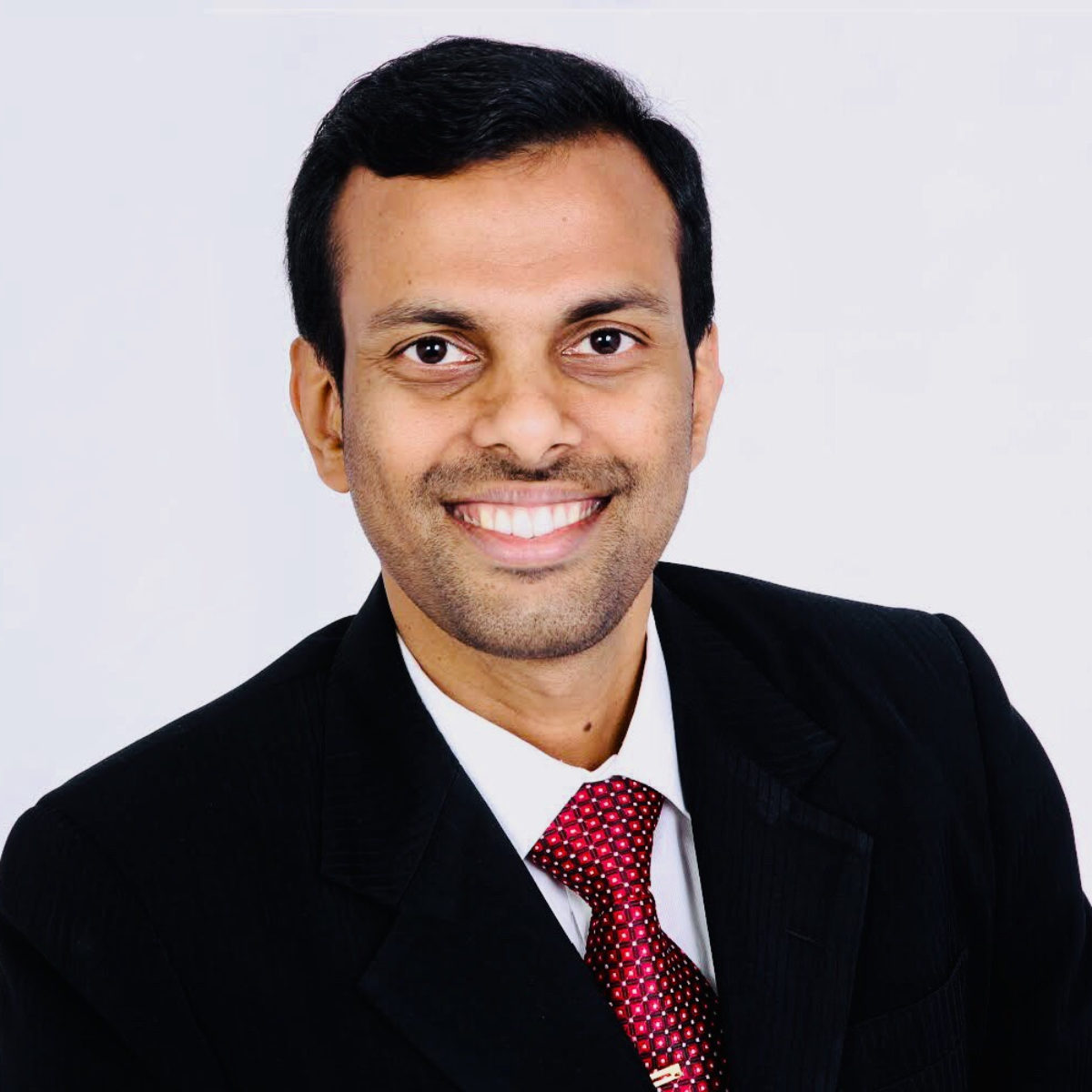}
\end{wrapfigure}\par
\textbf{Dr Mahendra Samarawickrama} (GAICD, MBA, SMIEEE, ACS(CP)) is the ICT Professional of the Year 2022 in the ACS Digital Disruptors Awards. He is a highly accomplished leader having an impressive track record of driving visions, technology innovations and transformation towards humanity, social justice, and sustainability. He is a founding director of the Centre for Ethical AI and the Centre for Sustainable AI. He supports the formation of organisational Environmental, Social, and Governance (ESG) strategy and drives ESG projects leveraging emerging technologies. He specialises in directing AI, Data Science and Customer Experience (CX)-focussed teams on building state-of-the-art capabilities. He is an author, inventor, mentor, advisor and regularly speaks at various technology forums, conferences and events worldwide. Many of his publications and frameworks related to AI governance and ethics are spotlighted in national and international forums.

As the Manager of the Data Science and Analytics team in the Australian Red Cross, he has developed an AI governance and strategy framework crucial to the business’ successful deployment of Data Science and AI capabilities to mobilise the power of humanity. He built the Volunteer Data Science and Analytics team from the ground up, supporting the Australian Red Cross’s strategic goals. He is supporting the business for personalised engagement of customers for disaster resilience in these demanding times of pandemic, natural disasters, and global conflicts. He is also a co-author of the IFRC data playbook and contributed to the data science and emerging technology chapter for AI governance, ethics, and literacy. In all these processes, he valued diversity, equity and inclusion. In recognition of this, his team became finalists in 1) the Diversity, Equity and Inclusion in Action Award in the 2021 IoT Awards, 2) the Best Use of Technology to Revolutionise CX Award in the 2021 Ashton Media CX Awards, 3) the Service Transformation for the Digital Consumer for Not-for-Profit/NGO in 2022 ACS Digital Disruptors Awards, and contributed to winning the CX Team of the Year Award in 2021 Ashton Media CX Awards. All of these awards are prestigious national awards.

He is an industry collaborator who actively leads technology innovation-and-transformation initiatives and partnerships toward humanity, social justice and sustainability. In this perspective, he is an Advisory Council Member in Harvard Business Review (HBR), an Expert in AI ethics and governance at Global AI Ethics Institute, an industry Mentor in the UNSW business school, a senior member of IEEE (SMIEEE), an honorary visiting scholar at the University of Technology Sydney (UTS), an Advisor for Data Science and Ai Association of Australia (DSAi), and a graduate member of the Australian Institute of Company Directors (GAICD).

He has recently established a YouTube channel and a Twitter channel to share his knowledge with the community. With a PhD in Computer Science and Masters degrees in Business Administration and Project Management, he brings the capacity to steer organisations through the complex, data-driven problems of our time.
\begin{figure}[!htbp]
    \centering
    \subfloat{{\includegraphics[width=3cm]{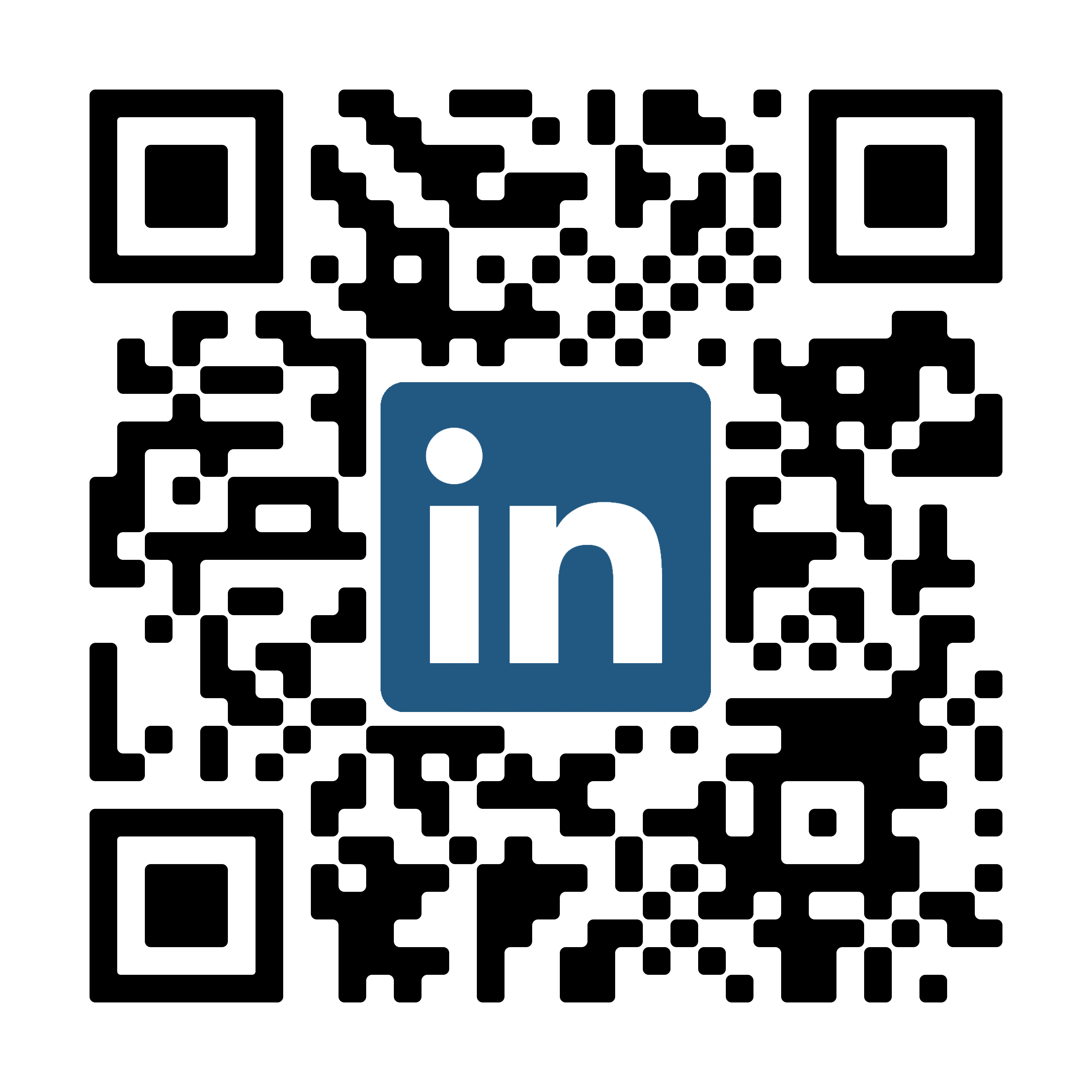} }}
    \qquad
    \subfloat{{\includegraphics[width=3cm]{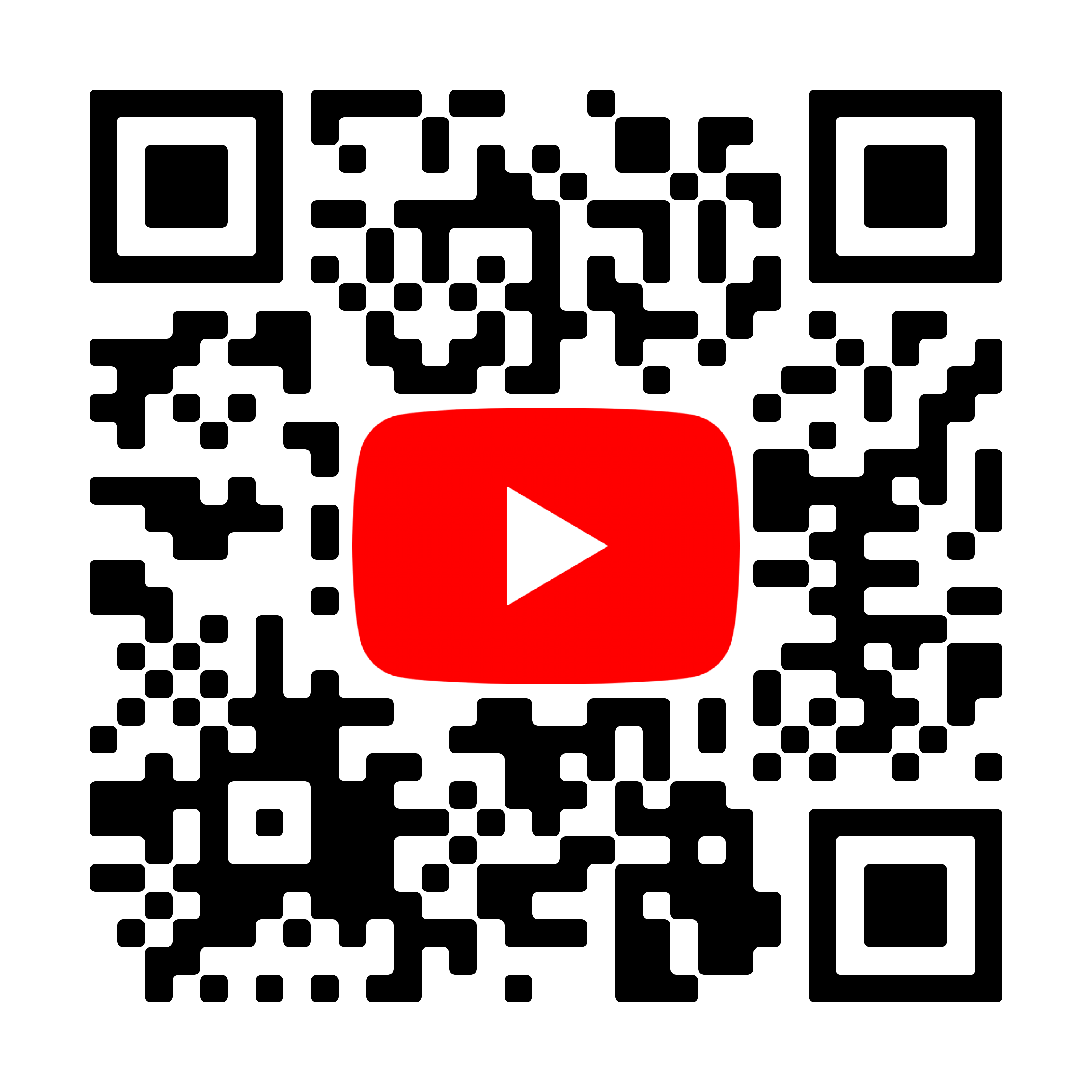} }}
    \qquad
    \subfloat{{\includegraphics[width=3cm]{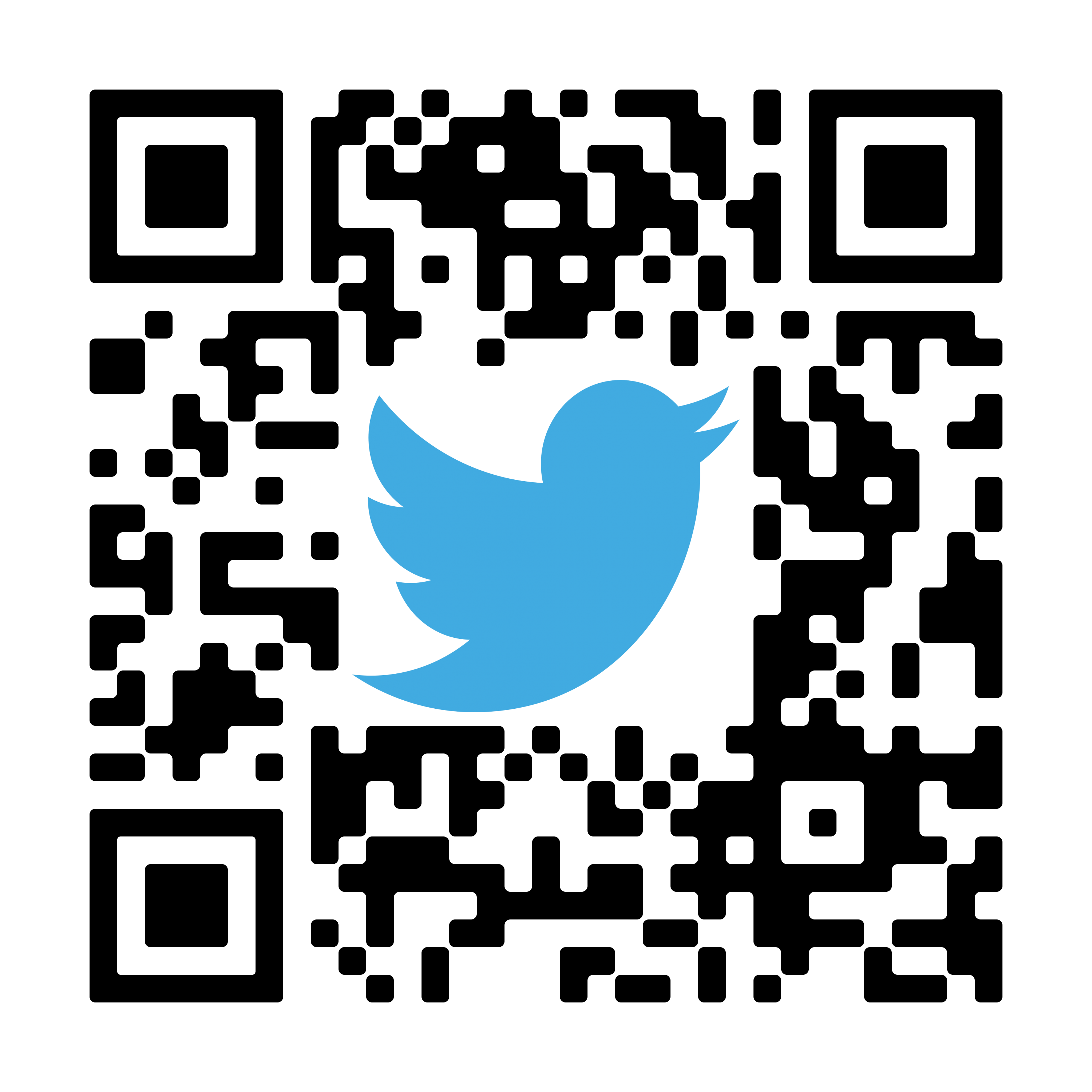} }}
\end{figure}

\end{document}